\documentclass[10pt,twoside,twocolumn,a4paper]{IEEEtran}
\usepackage{epsfig,bbm}
\usepackage{amsmath}
\usepackage{amsfonts}
\usepackage{cite}
\usepackage{graphicx}
\usepackage{xcolor}
\usepackage[ruled,linesnumbered]{algorithm2e}
\usepackage{multirow}
\usepackage{adjustbox}
\usepackage{gensymb}

\newtheorem{theorem}{\bf Proposition}

\newcommand{\A}{{\bf A}}

\newcommand{\x}{{\bf x}}

\newcommand{\I}{{\bf I}}

\newcommand{\tr}{{\tt tr}}

\newcommand{\C}{{\bf C}}
\newcommand{\wCxk}{\widehat{\bf C}^{k}}
\newcommand{\wCxkt}{\widehat{\bf C}^{k,t}}
\newcommand{\Cxkt}{{\bf C}^{k,t}}

\begin{document}

\title{Dynamic Independent Component/Vector Analysis: Time-Variant Linear Mixtures Separable by Time-Invariant Beamformers}

\author{{\bf Zbyn\v{e}k Koldovsk\'{y}$^1$, V\'aclav Kautsk\'y$^{1,2}$, Petr Tichavsk\'y$^3$, Jaroslav \v{C}mejla$^1$, and Ji\v{r}\'i M\'alek$^1$}
\vspace{0.1in} \\
$^1$Acoustic Signal Analysis and Processing Group, Faculty of Mechatronics, 
Informatics, and Interdisciplinary\\ Studies,
Technical University of Liberec, Studentsk\'a 2, 461 17
Liberec, Czech Republic. \\E-mail:
zbynek.koldovsky@tul.cz, fax:+420-485-353112, tel:+420-485-353534
\\
$^2$Faculty of Nuclear Sciences and Physical Engineering,
	Czech Technical University in Prague, Czech Republic. 
\\
$^3$The Czech Academy of Sciences, Institute of Information Theory and 
Automation,\\ Pod
vod\'{a}renskou v\v{e}\v{z}\'{\i} 4, P.O.~Box 18, 182 08 Praha 8,
Czech Republic. E-mail: tichavsk@utia.cas.cz,
fax:+420-2-868-90300, tel. +420-2-66052292
}

%
%
%
%


\maketitle

\footnotetext{This work was supported by 
The Czech Science Foundation through Project No.~20-17720S, and by the department of the Navy, Office of Naval Research Global, through Project No.~N62909-19-1-2105. Matlab implementations of FastDIVA and of the example presented in Section~I are available at {\tt https://asap.ite.tul.cz/downloads/ice/}.}

\begin{abstract}
A novel extension of Independent Component and Independent Vector Analysis for blind extraction/separation of one or several sources from time-varying mixtures is proposed. The mixtures are assumed to be separable source-by-source in series or in parallel based on a recently proposed mixing model that allows for the movements of the desired source while the separating beamformer is time-invariant. The popular FastICA algorithm is extended for these mixtures in one-unit, symmetric and block-deflation variants. The algorithms are derived within a unified framework so that they are applicable in the real-valued as well as complex-valued domains, and jointly to several mixtures, similar to Independent Vector Analysis. Performance analysis of the one-unit algorithm is provided; it shows its asymptotic efficiency under the given mixing and statistical models. Numerical simulations corroborate the validity of the analysis, confirm the usefulness of the algorithms in separation of moving sources, and show the superior speed of convergence and ability to separate super-Gaussian as well as sub-Gaussian signals. 
\end{abstract}

{\keywords Blind Source Separation, Blind 
Source Extraction, Independent Component Analysis, Independent Vector Analysis, Dynamic Models, Moving Sources}

\section{Introduction}\label{section:introduction}
Independent Component Analysis (ICA) is a popular method proposed for 
Blind Source Separation (BSS) \cite{herault1987, comon1994, cardoso1998}. 
Signals observed on $d$ sensors are assumed to be linear mixtures of $d$ ``original'' signals, which are mutually independent in the statistical sense. The linear mixing model is given by
\begin{equation}\label{eq:modelstaticICA}
\x^n = \A^n {\bf s}^n,
\end{equation} 
where $n=1,\dots,N$ is the sample index, $\x^n$ is a $d\times 1$ vector of the observed mixed signals at time $n$; $\A^n$ is a 
$d\times d$ non-singular mixing matrix; and ${\bf s}^n$ is a $d\times 1$ vector of the original independent signals. Since $\A^n$ are square and non-singular, the model is referred to as {\em determined}. We speak about the {\em static} mixing case when $\A^n$ is constant over $n$. ICA can be formulated as to estimate $(\A^n)^{-1}$ through finding square de-mixing matrices ${\bf W}^n$ such that the signals ${\bf W}^n\x^n$ are as independent as possible. It is the indeterminacy of BSS (as well as of ICA) that the order and scales of ${\bf s}^n$ cannot be retrieved without additional information.

The determined static formulation has become very popular mainly due to its mathematical tractability and wide applicability. The problem has been deeply studied and, currently, ICA and its extension to joint separation of several mixtures (data sets) such as Independent Vector Analysis (IVA), have matured to a large extent \cite{lee1998,hyvarinen2001,cichocki2002,comon2010handbook,adali2014b} . 
For most recent contributions to the area see, e.g., \cite{kitamura2018,brendel2020}.

In many applications, however, it is necessary to consider the time-variant mixing model, which we will refer to as {\em dynamic}.
For example, in audio or biomedical applications it happens that the mixing environment is changing in time, sources are moving, some new sources can appear randomly in time and some other may disappear. There is therefore a need to estimate the mixing/de-mixing matrix in an adaptive manner, respecting the dynamic nature of the data. 

The determined mixing model \eqref{eq:modelstaticICA} with $\A^n$ dependent on $n$ can capture a very wide class of dynamic mixtures. However, the lack of information (the number of samples $N$ is proportional to the number of unknown parameters $Nd^2$) and the random order of the separated signals at any time instant pose crucial problems. Current extensions of ICA and IVA towards the dynamic model are therefore based on more or less strictly formulated assumptions that the changes of the mixing parameters are somewhat slow and continuous. A standard way is that estimation methods for the static case are converted into adaptive algorithms 
along the lines of the least-mean-squares (LMS) or recursive-least-squares (RLS) algorithms \cite{haykin2010}. To this end, various sequential \cite{welling2004}, recursive \cite{taniguchi2014,hsu2016}, Bayesian \cite{chien2013} or other online approaches have been proposed. 
Particularly popular adaptive methods are based on the Natural Gradient algorithm \cite{amari1996,laheld1996}; see, e.g., \cite{mukai2005,nesta2011taslp,akhtar2012,khan2015,hsu2016}.
In biomedical application, an Online Recursive Independent Component Analysis (ORICA) was proposed in \cite{hsu2014,hsu2016}. 
The latter paper is remarkable because it presents a real-world application of the algorithm in high-density (64 channel) EEG.

The approach that we present here is conceptually different. Basically, it is off-line, despite it allows to handle time-varying mixtures to certain extent. It comes from the recently proposed blind source extraction (BSE) model denoted as CSV (Constant Separating Vector) where the mixing parameters related to the source of interest (SOI) can be varying in time while the de-mixing parameters are time-invariant \cite{koldovsky2019icassp,jansky2020}. CSV allows for the SOI movements throughout the exposed data. The Cram\'er-Rao analysis has been done in \cite{kautsky2020CRLB}. It points to appealing properties of CSV in terms of the achievable extraction accuracy compared to the sequentially applied ICA.
On-line version of the proposed approach is possible as well, because we can think about allowing the ``constant" separating vector to be progressively updated in time.

This paper brings two major contributions. First, we extend CSV to separation of more than one source, by which we introduce so-called {\em CSV-separable} mixtures. Briefly, the CSV-separable mixtures are defined as such that can be separated source-by-source in series or in parallel based on the CSV. In fact, the formulation of ICA/IVA on CSV-separable mixtures is a natural extension of the static ICA/IVA to the special class of dynamic mixtures. It provides a novel tool for off-line exploratory data analysis and is also useful in online data processing, as we demonstrate in the experimental section. Second, we propose the FastDIVA algorithm (Fast Dynamic Independent Vector Analysis) as a new method for ICA/IVA on CSV-separable mixtures. In fact, FastDIVA is a successor of the famous FastICA \cite{hyvarinen1999} and FastIVA \cite{lee2007fast} as it involves these methods as special cases and is proposed in three variants: one-unit, symmetric and block-deflation. To motivate, we provide the following example.

\begin{figure}[htbp]
    \centering
    \includegraphics[width=0.49\linewidth]{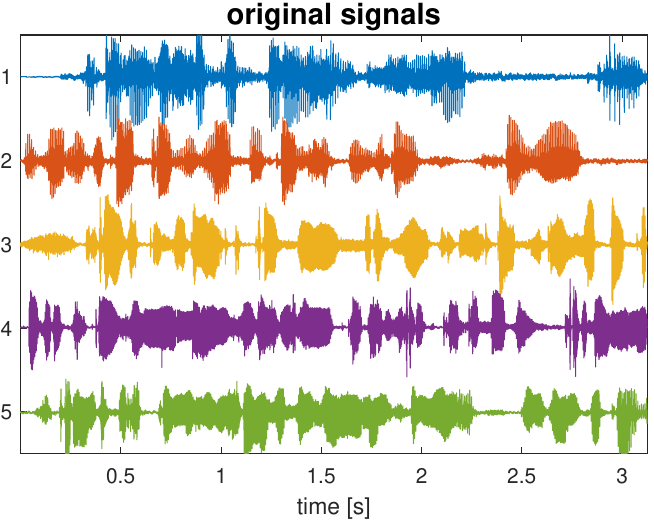}
    \includegraphics[width=0.49\linewidth]{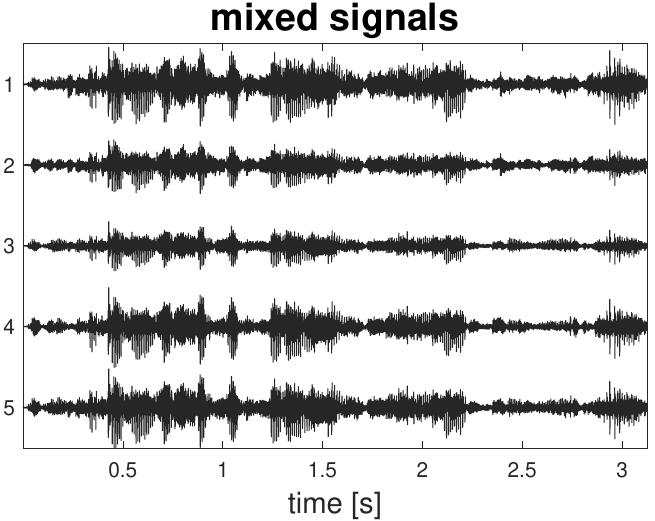}
    \caption{LEFT: Five independent speech signals, each $50,000$ samples long, sampled at $16$~kHz. RIGHT: Instantaneous mixture of the signals where the first signal is linearly moving while signals $2$ through $5$ are static. Signal $1$ is dominating the mixture (multiplied by factor $5$).}
    \label{fig:speechsignals}
\end{figure}

Consider five speech signals shown in Fig.~\ref{fig:speechsignals} (left). Their instantaneous\footnote{Note that this mixture is not convolutive as is typical to real-world acoustic signal mixing; we consider the simpler instantaneous case for  demonstration purposes.} mixture, shown in Fig.~\ref{fig:speechsignals} (right), is generated so that signals $2$ through $5$ are static, mixed into $5$ channels with fixed random  mixing coefficients, while signal $1$ (i.e., its virtual source) is moving: The first column of ${\bf A}^n$, denoted as ${\bf a}^n$, is linearly progressing from ${\bf a}$ to ${\bf b}$ according to ${\bf a}^n=(N-n+1)/N{\bf a}+(n-1)/N{\bf b}$; ${\bf a}$ and ${\bf b}$ are random column vectors that make an angle of $20^{\circ}$. The other columns of ${\bf A}^n$ are constant over $n$. Signal $1$ is amplified by factor $5$ in order to accentuate it in the mixture. 

\begin{figure}[htbp]
    \centering
    \includegraphics[width=0.8\linewidth]{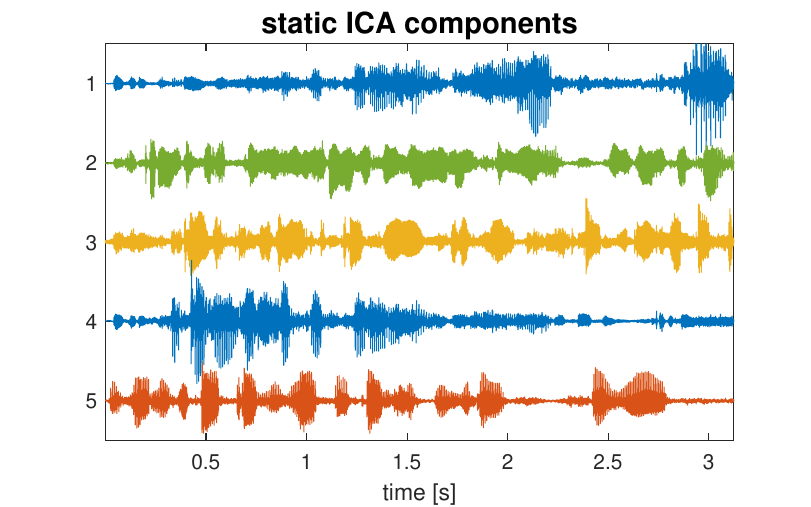}
    \caption{Independent components extracted from the signal mixture shown in the right part of Fig.~\ref{fig:speechsignals} by symmetric FastICA.}
    \label{fig:efica}
\end{figure}

Fig.~\ref{fig:efica} shows typical components obtained by a conventional ICA algorithm (symmetric FastICA \cite{hyvarinen1999})  when applied to this mixture. The order of components is random, which is due to the inherent ambiguity of BSS. By visual inspection, components $2$ and $5$ correspond to the original signals $5$ and $2$, respectively, up to scales and signs. Component $3$ corresponds to the original signal $3$ up to a certain more significant residual interference. Components $4$ and $1$ consist of the beginning and end parts of  the original signal $1$, respectively. This is caused by the movement of the corresponding (virtual) source. The original signal $4$ is not extracted as a separated component; it appears as a residual within component $4$. Note that this situation cannot be improved by extracting one more component because the static ICA (de-)mixing model assumes square (de-)mixing matrix.

\begin{figure}[htbp]
    \centering
    \includegraphics[width=0.8\linewidth]{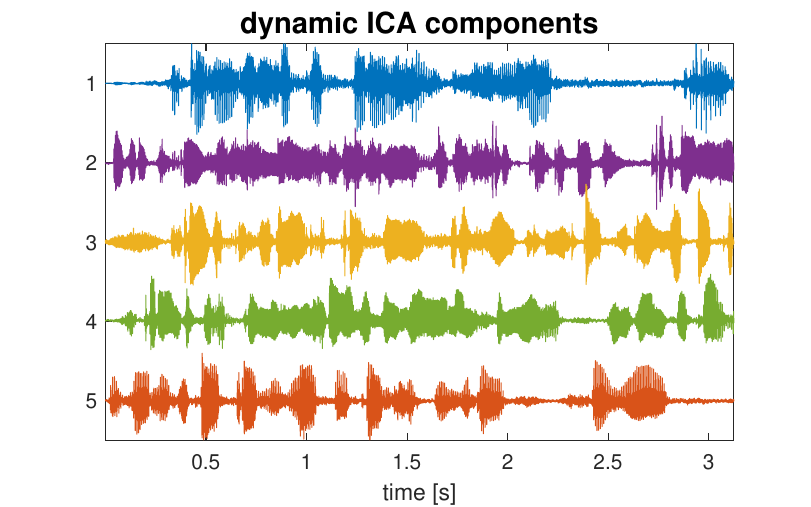}
    \caption{Independent components separated from the dynamic mixture in Fig.~\ref{fig:speechsignals} (right) by block-deflation FastDIVA set to $5$ blocks, each of length $10^4$ samples.}
    \label{fig:fastdiva}
\end{figure}

Fig.~\ref{fig:fastdiva} shows components that have been separated by block-deflation FastDIVA assuming CSV-separable mixing model with $5$ blocks. The components correspond with the original signals up to a random order, which is 1, 4, 3, 5, 2, and a reasonable statistical error. Not only does the algorithm extract the moving signal as one component, that is, without the need for collecting it from several components whose order is unknown. It also separates original signal $4$ with a high degree of precision, as compared to symmetric FastICA. 

This paper is organized as follows. The CSV mixing model is revised and extended to the CSV-separable mixtures in Section~II. The FastDIVA algorithm is proposed in Section~\ref{section:fastdiva} and its one-unit version is analyzed in Section~\ref{section:analysis}. Numerical experiments and comparisons with FastDIVA in off-line and on-line tests are provided in Section~V; and Section~VI concludes the article. 

\subsection*{Nomenclature and conventions} 
Plain, bold, and bold capital letters denote scalars, vectors, and matrices, respectively. Upper index $\cdot^T$, $\cdot^H$, or $\cdot^*$ denotes, respectively, transposition, conjugate transpose, or complex conjugate. The Matlab convention for matrix/vector concatenation will be used, e.g., $[1;\,{\bf g}]=[1,\, {\bf g}^T]^T$. 
The statistical models of signals considered in this paper assume that each sample is independently drawn from a distribution; inter-sample dependencies are not modeled. Therefore, we use symbolic notation where samples having the same distribution are represented by random (vector) variables. ${\rm E}[\cdot]$ stands for the expectation value of the argument, and $\hat{\rm E}[\cdot]$ is the average value of the argument taken over all of its available samples. 
The letters $k$, $t$, and $i$ are used as integer indices of dataset, block, and source, respectively; index omission will always be announced in the text. $\{\cdot\}_k$ is a short notation of the argument with all values of index $k$, e.g., $\{{\bf w}^k\}_k$ means ${\bf w}^1,\dots,{\bf w}^K$. The average value of $a_t$ taken over all available blocks, i.e., $\frac{1}{T}\sum_{t=1}^T a_t$, is denoted by $\left< a_t\right>_t$.

We will consider complex-valued signals and parameters; however, the conclusions of this work are valid for the real-valued case as well.

\section{Problem Formulation}\label{section:separable}
For practical reasons, we turn from \eqref{eq:modelstaticICA} to mixtures that are block-wise static and, also, extend our considerations to multiple datasets as in IVA and in other joint BSS problems \cite{lee1997,adali2015,lahat2016,chen2016,weiss2018}. 

Let $N$ samples of signals be observed through $d$ sensors in $K$ datasets, and let the samples be divided into $T\geq 1$ time-intervals called blocks. For the sake of simplicity, let the blocks have the same length $N_b$, and $N=T\cdot N_b$. From now on, we will consider the block-wise varying mixing model
\begin{equation}\label{eq:mixingmodel}
    {\bf x}^{k,t}={\bf A}^{k,t}{\bf s}^{k,t},
\end{equation}
where $k=1,\dots,K$ is the dataset index; $t=1,\dots,T$ is the block index; ${\bf A}^{k,t}$ is a $d\times d$ non-singular mixing matrix; and by ${\bf s}^{k,t}=[s^{k,t}_1,\dots,s^{k,t}_d]^T$ we denote {\em independent} random variables representing unknown original signals. Without any loss of generality, let all the signals have zero mean values; samples of signals within the blocks are assumed identically and independently distributed (i.i.d.). By taking into account the ambiguities, the BSS task can be, in general, formulated as follows.

\begin{quote}
Find de-mixing matrices ${\bf W}^{k,t}$ such that ${\bf W}^{k,t}{\bf x}^{k,t}$ are equal to ${\bf s}^{k,t}$ up to their original scales and phase. The order of the separated signals can be different from the original one; however, it is desirable for it to be the same in all datasets and blocks.
\end{quote}

For $T=1$, we have the {\em static} case considered by the conventional ICA and IVA. In ICA, the datasets are separated independently; this approach, however, brings random permutation of separated signals in the datasets, the permutation problem \cite{sawada2004sap}. In IVA, components are separated as ``vectors'' where the $i$th vector component is defined as ${\bf s}_i^t=[s^{1,t}_i,\dots,s^{K,t}_i]^T$, $i=1,\dots,d$ \cite{kim2006}.

We are mainly interested in the dynamic case of $T>1$, where the mixing parameters (matrices) can be varying from block to block.
ICA and IVA can be used when $T>1$ by applying them separately on blocks. However, this approach does not guarantee the same order, i.e., continuity of the separated signals over the blocks, a phenomenon caused by the uncertainty of signal order similar to the permutation problem; we refer to it as {\em the discontinuity problem}. Also, there are too many parameters to be estimated, which potentially leads to a loss in separation accuracy.

What we basically do, in this paper, is applying a deflation or symmetric manner of blind source separation as in deflation or symmetric FastICA when $T=1$ \cite{hyvarinen1999}. It means that we wish to separate the signal components one by one or in parallel. Therefore, the primary problem to be solved is the blind extraction of one component. For $T=1$, this is solved through so-called  Independent Component or Independent Vector Extraction (ICE/IVE). 


In ICE/IVE, it is reflected that if only the SOI should be extracted, only the corresponding column of ${\bf A}^{k,t}$ and the corresponding row of $({\bf A}^{k,t})^{-1}$ need to be taken into account in the mixing model parameterization. The other columns of ${\bf A}^{k,t}$ need not be estimated, only their corresponding subspace should be identified. The parameterization chosen in \cite{koldovsky2019TSP} ensures this.

Owing to the indeterminacy of order in BSS\footnote{In fact, any knowledge about the SOI (e.g., an initial guess) must be available to determine it; see Section~II.B in \cite{koldovsky2019TSP}.}, we can assume that the SOI corresponds to $s_1^{k,t}$. Hence, according to \cite{koldovsky2019TSP}, ${\bf A}^{k,t}$ in \eqref{eq:mixingmodel} can be parameterized by
\begin{equation}\label{eq:mixingmatrixIVE}
{\bf A}_{\rm BSE}^{k,t}=
 \begin{pmatrix}
 {\bf a}^{k,t} & {\bf Q}^{k,t}
\end{pmatrix}  =
 \begin{pmatrix}
  \gamma^{k,t} & ({\bf h}^{k,t})^H\\
   {\bf g}^{k,t} &  \frac{1}{\gamma^{k,t}}({\bf g}^{k,t}({\bf h}^{k,t})^H-\I_{d-1})
    \end{pmatrix}
\end{equation}
where ${\bf a}^{k,t}=[\gamma^{k,t};{\bf g}^{k,t}]$ is called the {\em mixing vector} corresponding to the first column of ${\bf A}^{k,t}$; $\I_{d}$ denotes the $d\times d$ identity matrix; $({\bf w}^{k,t})^H$ denotes the first row of ${\bf W}^{k,t}_{\rm BSE}=({\bf A}^{k,t}_{\rm BSE})^{-1}$; ${\bf w}^{k,t}$ stands for the beamformer on which output is the extracted signal, i.e., $s_1^{k,1}=({\bf w}^{k,t})^H{\bf x}^{k,t}$; we will call it the {\em separating vector}. It holds that
\begin{equation}\label{eq:demixingmatrixIVE}
    {\bf W}_{\rm BSE}^{k,t} =
     \begin{pmatrix}
     ({\bf w}^{k,t})^H\\
     {\bf B}^{k,t}
     \end{pmatrix}  =
     \begin{pmatrix}
     (\beta^{k,t})^* & ({\bf h}^{k,t})^H\\
     {\bf g}^{k,t} & -\gamma^{k,t} \I_{d-1}
     \end{pmatrix},
\end{equation}
where ${\bf w}^{k,t}=[\beta^{k,t};{\bf h}^{k,t}]$, ${\bf B}^{k,t}=[{\bf g}^{k,t},\,-\gamma^{k,t} \I_{d-1}]$ satisfies the condition ${\bf B}^{k,t}{\bf a}^{k,t}={\bf 0}$ (the blocking matrix \cite{gannot2001}). 
Since ${\bf W}_{\rm BSE}^{k,t}{\bf A}_{\rm BSE}^{k,t}=\I_d$, ${\bf a}^{k,t}$ and ${\bf w}^{k,t}$ are linked through the so-called {\em distortionless constraint} $({\bf w}^{k,t})^H{\bf a}^{k,t} = 1$, which can  also be written as
\begin{equation}\label{eq:distortionlessconstraint}
(\beta^{k,t})^*\gamma^{k,t}=1-({\bf h}^{k,t})^H{\bf g}^{k,t}.
\end{equation}
The subspace of the other signals, referred to as {\em background}, is generated by ${\bf z}^{k,t}={\bf B}^{k,t}{\bf x}^{k,t}$. The fact that ${\bf B}^{k,t}{\bf a}^{k,t}={\bf 0}$ guarantees that ${\bf z}^{k,t}$ span the same subspace as $s_2^{k,t},\dots,s_d^{k,t}$.

The approach where ICE/IVE is applied separately to each block when $T>1$ has the same drawbacks as mentioned above. In \cite{koldovsky2019icassp,kautsky2020CRLB}, we have distinguished two simplifications, or two particular separation models.
\begin{itemize}
\item Constant mixing vector (CMV)
\begin{equation}\label{eq:CMV}
{\bf a}^{k,1}={\bf a}^{k,2}=\ldots ={\bf a}^{k,T}={\bf a}^{k}
\end{equation}
\item Constant separating vector (CSV)
\begin{equation}\label{eq:CSV}
{\bf w}^{k,1}={\bf w}^{k,2}=\ldots ={\bf w}^{k,T}={\bf w}^{k}
\end{equation}
\end{itemize}
Each of the two models ensures that the components in different blocks are not permuted randomly, thus, the discontinuity problem is avoided. The former model might be more suitable when the sources are not moving, but the background is non-stationary, or that there is low signal-to-interference and noise ratio. The latter models appears to be useful when the SOI is moving, which is of a greater interest. Therefore, we deal with the CSV model, in this paper, and generalize it to separation of $r\geq 1$ sources. 

We introduce the notion of {\em CSV-separable mixtures} through the following conditions:
\begin{enumerate}
    \item[(C1)] All $r$ sources to be separated obey CSV, which means that the first $r$ rows of $({\bf A}^{k,1})^{-1},\dots,({\bf A}^{k,T})^{-1}$ in \eqref{eq:mixingmodel} are constant over $t$.
    \item[(C2)] For each $i=1,\dots,r$, the $i$th source obeys CSV in a reduced mixture where sources $1,\dots,i-1$ have been subtracted.
\end{enumerate}
Some properties readily follow. For $r=d$, all rows of the inverse matrices are assumed constant in (C1), which means that the mixtures obeying (C1) are static when $r=d$. For $r=1$, (C1) and (C2) coincide with the CSV model for one source. 

Validity of the conditions (C1) and (C2) has to be assumed. Their usefulness was already shown in Section~I and will be supported by additional examples in Section~V.
In Section~\ref{section:BSS}, we propose the symmetric and block-deflation separation schemes\footnote{The symmetric, deflation and block-deflation separation schemes are here applied together with the one-unit FastDIVA algorithm, however, they can be applied with other BSE algorithms.}, which can be used to separate $r$ sources from mixtures obeying (C1) and (C2), respectively. 


\section{Proposed Algorithm}\label{section:fastdiva}
The detailed derivations of one-unit, symmetric and block-deflation FastDIVA are provided in this Section. We begin with the one-unit variant, which solves the BSE problem based on the CSV mixing model.

\subsection{Statistical model}
To simplify notation, for now, we will omit the subscript ``$1$'' in ${\bf s}_1^t$, i.e., ${\bf s}^t=[s^{1,t},\dots,s^{K,t}]^T$. Let the probability density function (pdf) of ${\bf s}^t$ be $p({\bf s}^t)$. Note that this pdf is, in general, dependent on $t$; we do not write this explicitly, for simplicity. 
Next, let $p_{{\bf z}^{k,t}}({\bf z}^{k,t})$ denote the pdf of ${\bf z}^{k,t}$. Although there can  also be dependencies between background signals from different datasets, we neglect them to simplify the statistical model of the background\footnote{This simplification typically brings a suboptimal performance of BSE as compared to BSS \cite{kautsky2020CRLB,kautsky2017}.}. 

Considering the structure of the de-mixing matrix \eqref{eq:demixingmatrixIVE} with the CSV assumption \eqref{eq:CSV}, using the independence between the SOI and the background, and taking into account the fact that samples are independently distributed, we get the joint pdf for one sample of the observed signals in the $t$th block in the form
\begin{multline}\label{eq:jointmixedpdf}
   p_{{\bf x}^{k,t}}(\{{\bf x}^{k,t}\}_k) = p(\{({\bf w}^k)^H{\bf x}^{k,t}\}_k)\times\\ \prod_{k=1}^K p_{{\bf z}^{k,t}}({\bf B}^{k,t}{\bf x}^{k,t}) |\det {\bf W}^{k,t}_{\rm BSE}|^2.
\end{multline}
Note that the square of the absolute value of determinant is necessary due to the transformation of densities of the complex-valued random variables (the exponent equals one in the real-valued case). The determinant can be expressed by using Eq. (15) in \cite{koldovsky2019TSP}, which gives, together with \eqref{eq:CSV}, $|\det {\bf W}^{k,t}_{\rm BSE}|^2=|\gamma^{k,t}|^{2(d-2)}$.
The pdf of all $N$ samples is equal to $\prod_{t=1}^T p_{{\bf x}^{k,t}}(\{{\bf x}^{k,t}\}_k)^{N_b}$, so the log-likelihood function divided by $N$ can be expressed as
\begin{multline}\label{eq:loglikelihood}
    \mathcal{L}\left(\{{\bf w}^k,{\bf a}^{k,t}\}_{k,t}\right) =\Big<\hat{\rm E}\left[\log p\left(\left\{({\bf w}^k)^H{\bf x}^{k,t}\right\}_k\right)\right] \\ +\sum_{k=1}^{K} \hat{\rm E}\left[p({\bf B}^{k,t}{\bf x}^{k,t})\right] 
    + (d-2)\sum_{k=1}^{K}\log |\gamma^{k,t}|^2\Big>_t.
\end{multline}

\subsection{Contrast function}
Finding the appropriate maximum of \eqref{eq:loglikelihood} provides the maximum likelihood estimate of the parameter vectors. However, \eqref{eq:loglikelihood} must be replaced by a valid contrast function because of the unknown pdfs $p({\bf s}^t)$ and $p_{{\bf z}^{k,t}}({\bf z}^{k,t})$, which have to be replaced by suitable model densities. In the static case, the model pdfs of the SOI can be scaled to unit variance since there is the scaling ambiguity \cite{hyvarinen1999}. However, in the dynamic case, the variance of signals can be changing from block to block which must be taken into account. 
Therefore, the appropriate surrogate for $p(\cdot)$ is\footnote{Note that the square power in \eqref{eq:modeldensity} is necessary due to considering the complex-valued problem; it would equal one in the real-valued case.} \cite{pham2001b,pham2006}
\begin{equation}\label{eq:modeldensity}
    p({\bf s}^t) \approx f\left(\left\{\frac{s^{k,t}}{\hat\sigma^{k,t}}\right\}_k\right)\left(\prod_{k=1}^K\hat\sigma^{k,t}\right)^{-2},
\end{equation}
where $f(\cdot)$ should be a suitable normalized non-Gaussian pdf, and $(\hat\sigma^{k,t})^2$ is the sample-based variance of the estimate of $s^{k,t}$. It holds that 
\begin{equation}\label{eq:sigma}
   \hat\sigma^{k,t}=\sqrt{({\bf w}^k)^H\widehat{\bf C}^{k,t}{\bf w}^k},
\end{equation}
where $\widehat{\bf C}^{k,t}=\hat{\rm E}[{\bf x}^{k,t}({\bf x}^{k,t})^H]$ is the sample-based covariance matrix of ${\bf x}^{k,t}$; $\hat\sigma^{k,t}$ is, in fact, a function of ${\bf w}^k$.

Note that $f(\cdot)$ could be dependent on $t$. However, since there is usually little information about the true pdf, we simplify our considerations by assuming that $f(\cdot)$ is independent of $t$.

The unknown $p_{{\bf z}^{k,t}}({\bf z}^{k,t})$ can be replaced by the zero mean circular\footnote{Noncircular Gaussian pdf could be considered as well, especially, if the background signals are assumed to involve noncircular sources. In Appendix A, it will be shown that the assumption of circularity causes that the Hessian matrix ${\bf H}_1$, defined later in \eqref{eq:H1}, has rank 1, which significantly simplifies the Newton-Raphson update given by \eqref{eq:newtonraphsonupdate}.} Gaussian pdf $\mathcal{CN}(0,{\bf C}_{\bf z}^{k,t})$, where
${\bf C}_{\bf z}^{k,t}={\rm E}[{\bf z}^{k,t}({\bf z}^{k,t})^H]$ is the covariance matrix of the background signals; see, e.g., \cite{koldovsky2019TSP} for the justification of this choice. ${\bf C}_{\bf z}^{k,t}$ is an unknown nuisance parameter, which will   later be replaced by its sample-based estimate. By putting the model densities into \eqref{eq:loglikelihood}, a practical contrast function for estimating the model parameters takes on the form
\begin{multline}\label{eq:contastCSV}
    \mathcal{C}\left(\{{\bf w}^k,{\bf a}^{k,t}\}_{k,t}\right) =\Bigg<\hat{\rm E}\left[\log f\left(\left\{\frac{\hat{s}^{k,t}}{\hat\sigma^{k,t}}\right\}_k\right)\right]  \\ -\sum_{k=1}^{K}\log(\hat\sigma^{k,t})^2 -\sum_{k=1}^{K} \hat{\rm E}\left[(\hat{\bf z}^{k,t})^H({\bf C}_{\bf z}^{k,t})^{-1}\hat{\bf z}^{k,t}\right]  \\ 
    + (d-2)\sum_{k=1}^{K}\log |\gamma^{k,t}|^2\Bigg>_t + \text{const.},
\end{multline}
where $\hat{s}^{k,t}=({\bf w}^k)^H{\bf x}^{k,t}$, and 
$\hat{\bf z}^{k,t}={\bf B}^{k,t}{\bf x}^{k,t}$. The remaining constant term is independent of the mixing model parameters.
For $K=1$ and $T=1$, the indices $k$ and $t$ can be omitted, and \eqref{eq:contastCSV} is simplified to\footnote{The reader can compare \eqref{eq:contastICE} with Equation 19 in \cite{koldovsky2019TSP}. The contrast functions differ in that \eqref{eq:contastICE} involves $\hat\sigma^2$; therefore, it contains the normalization inside the argument of $f(\cdot)$ and an additional second term.} 
\begin{multline}\label{eq:contastICE}
    \mathcal{C}^{1,1}\left({\bf w},{\bf a}\right) =\hat{\rm E}\left[\log f\left(\frac{\hat{s}}{\hat\sigma}\right)\right]  -\log\hat\sigma^2 -\hat{\rm E}\left[\hat{\bf z}^H{\bf C}_{\bf z}^{-1}\hat{\bf z}\right]  \\ 
    + (d-2)\log |\gamma|^2 + \text{const.}
\end{multline}

\subsection{Orthogonal constraints}
The above contrast functions can have many spurious extremes.
It may occur that the parameter vectors ${\bf a}^{k,t}$, $t=1,\dots,T$, and ${\bf w}^{k}$ do not correspond to the same signal. 
Therefore, a reliable link between the separating and mixing vectors has to be established. To this end, the orthogonal constraint (OGC) appears to be convenient. Since $s^{k,t}$ and ${\bf z}^{k,t}$ are independent and, therefore, also uncorrelated, the OGC requires that subspace generated by samples of $\hat{s}^{k,t}$ is orthogonal to the subspace of $\hat{\bf z}^{k,t}$. Also, \eqref{eq:distortionlessconstraint} must be satisfied. The mixing vectors are then linked  with the separating vector through \cite{koldovsky2019TSP}
\begin{equation}\label{eq:orthogonalconstraint}
	{\bf a}^{k,t}=\frac{\wCxkt{\bf w}^k}{({\bf w}^k)^H\wCxkt{\bf w}^k}.
\end{equation}
Equivalently, ${\bf w}^k$ can be expressed as the dependent variable as
\begin{equation}\label{eq:orthogonalconstraint2}
	{\bf w}^{k}=\frac{(\wCxkt)^{-1}{\bf a}^{k,t}}{({\bf a}^{k,t})^H(\wCxkt)^{-1}{\bf a}^{k,t}}.
\end{equation}

\subsection{Relationship to optimum beamformers}\label{section:LCMP}
The analytic expression \eqref{eq:orthogonalconstraint2} corresponds to the minimum power distortionless beamformer (MPDR) steered in the direction determined by the mixing vector ${\bf a}^{k,t}$ when the covariance of data is given by $\wCxkt$. MPDR is an optimum beamformer known in array processing theory as the solution of  \cite{vantrees2002}
\begin{equation}\label{eq:MPDR}
	{\bf w}^{k}=\arg\min_{\bf w} {\bf w}^H\wCxk{\bf w} \quad \text{w.r.t.} \quad {\bf w}^H{\bf a}^{k,t}=1.
\end{equation}
The orthogonally constrained BSE algorithms can, in the static case of $T=1$, be viewed as blind MPDR beamformers seeking in the direction of ${\bf a}^{k,t}$, for a fixed $t$, such that the MPDR output is independent of the orthogonal (background) subspace \cite{koldovsky2017eusipco}. 

In the CSV model, \eqref{eq:orthogonalconstraint2} and, thus, \eqref{eq:MPDR} should be satisfied simultaneously for all $t=1,\dots,T$, which imposes $T$ conditions on one separating vector ${\bf w}^k$. It is therefore more practical to impose the OGC through \eqref{eq:orthogonalconstraint} rather than through \eqref{eq:orthogonalconstraint2} when $T>1$.

In order to interpret the block-independent separating vector in CSV, note that the true mixing and separating vectors satisfy \eqref{eq:orthogonalconstraint2} when $N_b\rightarrow+\infty$, that is, with $\wCxkt$ replaced by $\Cxkt$. Hence, the true parameter vectors satisfy 
\begin{equation}
	{\bf w}^k = \arg\min_{\bf w} {\bf w}^H\Cxkt{\bf w} \quad \text{w.r.t.} \quad {\bf w}^H{\bf a}^{k,t}=1
\end{equation}
for all $t=1,\dots,T$. It follows that they also obey
\begin{equation}\label{eq:LCMP}
	{\bf w}^k = \arg\min_{\bf w} {\bf w}^H{\bf R}^k{\bf w} \quad \text{w.r.t.} \quad {\bf w}^H{\boldsymbol{\Lambda}}^{k}={\bf 1},
\end{equation}
where ${\bf R}^k=\sum_{t=1}^T\Cxkt$, $\boldsymbol{\Lambda}^{k}=[{\bf a}^{k,1}\dots {\bf a}^{k,T}]$, and ${\bf 1}$ is the $T\times 1$ vector of ones. The solution of \eqref{eq:LCMP} is known as the linearly constrained minimum power beamformer (LCMP) \cite{vantrees2002}. We conclude the connection between CSV and LCMP as follows:
\begin{quote}
    For $N_b\rightarrow +\infty$, the CSV mixing model ensures that the LCMP beamformer steered in the directions given by the true mixing vectors (determining locations of the SOI during its movement)
${\bf a}^{k,1},\dots,{\bf a}^{k,T}$ exists such that it extracts the SOI from the mixed signals perfectly.
\end{quote} 

\subsection{Approximate Newton-Raphson algorithm}
The algorithm proposed here aims at finding a maximum of \eqref{eq:contastCSV} subject to the parameter vectors ${\bf w}^k$, $k=1,\dots,K$, under the OGC \eqref{eq:orthogonalconstraint}. 
For the sake of clarity, the contrast function to be maximized is
\begin{equation}\label{eq:contrastOG}
    \mathcal{C}_{\rm OG}\left(\left\{{\bf w}^k\right\}_{k}\right) =\mathcal{C}\left(\left\{{\bf w}^k,\frac{\wCxkt{\bf w}^k}{({\bf w}^k)^H\wCxkt{\bf w}^k}\right\}_{k,t}\right).
\end{equation}
We follow the complex-valued Newton-Raphson optimization approach using the Wirtinger calculus \cite{li2008}. This entails the computation of the gradient and the second-order derivatives of $\mathcal{C}_{\rm OG}$. 
To simplify the exposition, the derivations here will be done as if $T=1$ and $K=1$ (the indices $t$ and $k$ will be omitted); the result for $T\geq 1$ and $K\geq 1$ will readily follow. Thus, we now compute the derivatives of the four terms in  \eqref{eq:contastICE} when ${\bf a}=\frac{\widehat\C{\bf w}}{{\bf w}^H\widehat\C{\bf w}}$.

To compute the gradient, we use results from \cite{koldovsky2019TSP} and the following identities
\begin{align}
    \frac{\partial}{\partial {\bf w}^H}{\hat s}&=\frac{\partial}{\partial {\bf w}^H}{\bf w}^H{\bf x}={\bf x},\\
    \frac{\partial}{\partial {\bf w}^H}\frac{1}{\hat\sigma}&=
    \frac{\partial}{\partial {\bf w}^H}\frac{1}{\sqrt{{\bf w}^H\widehat{\bf C}{\bf w}}}=-\frac{\bf a}{2\hat\sigma},\\
    \frac{\partial}{\partial {\bf w}^H}\log\hat\sigma^2&= \frac{\partial}{\partial {\bf w}^H}\log\hat{\bf w}^H\widehat{\bf C}{\bf w}={\bf a}.\label{eq:difflogsigma}
\end{align}
The derivative of the first term in \eqref{eq:contastICE} reads
\begin{equation}\label{eq:grad1}
    \frac{\partial}{\partial {\bf w}^H}\hat{\rm E}\left[\log f\left(\frac{\hat{s}}{\hat\sigma}\right)\right]=-\hat{\rm E}\left[\phi\left(\frac{\hat{s}}{\hat\sigma}\right)\frac{\bf x}{\hat\sigma}\right]+\Re\{\hat\nu\}{\bf a},
\end{equation}
where $\hat\nu$ is the sample-based estimate of
\begin{equation}\label{eq:nu}
\nu = {\rm E}\left[\phi\left(\frac{{s}}{\sigma}\right)\frac{s}{\sigma}\right],
\end{equation}
$\Re\{\cdot\}$ denotes the real part of the argument, and
\begin{equation}\label{eq:score}
\phi(s) = -\frac{\partial}{\partial {s}^*} \log f(s)
\end{equation}
is the score function corresponding to the model density $f(\cdot)$. The derivative of the second term in \eqref{eq:contastICE} follows directly from \eqref{eq:difflogsigma}. The derivatives of the third and fourth terms are simplified to 
\begin{equation}\label{eq:grad3}
    \frac{\partial}{\partial {\bf w}^H} \left\{-\hat{\rm E}\left[\hat{\bf z}^H{\bf C}_{\bf z}^{-1}\hat{\bf z}\right]     + (d-2)\log |\gamma|^2\right\} = {\bf a},
\end{equation}
when ${\bf C}_{\bf z}$ is (after taking the derivative) replaced by $\widehat{\bf C}_{\bf z}$ as shown in Appendix~C in \cite{koldovsky2019TSP}. Hence, terms $2$ through $4$ in \eqref{eq:contastICE} do not contribute to the gradient as their derivatives finally boil down to zero\footnote{It follows that BSE methods based on maximizing the non-Gaussianity of the SOI \cite{huber1985,comon1994,hyvarinen1999}, in fact, inherently assume that the background is circular Gaussian with unknown covariance.}. The gradient of \eqref{eq:contrastOG} for $T=K=1$ is thus equal to \eqref{eq:grad1}, i.e., 
\begin{equation}\label{eq:grad}
    \frac{\partial}{\partial {\bf w}^H}\mathcal{C}_{\rm OG}({\bf w})=\Re\{\hat\nu\}{\bf a}-\hat{\rm E}\left[\phi\left(\frac{\hat{s}}{\hat\sigma}\right)\frac{\bf x}{\hat\sigma}\right].
\end{equation}

Now, consider $N\rightarrow +\infty$ and ${\bf w}$ being the true separating vector; if this is the case, \eqref{eq:grad} is equal to
\begin{equation}
   \frac{\partial}{\partial {\bf w}^H}\mathcal{C}_{\rm OG}({\bf w})=( \Re\{\nu\}-\nu){\bf a}.
\end{equation}
It follows that the true separating vector is the stationary point of $\mathcal{C}_{\rm OG}({\bf w})$ only if $\Re\{\nu\}=\nu$. If $f(\cdot)=p(\cdot)$ then $\nu=1$, and the condition $\Re\{\nu\}=\nu$ is satisfied. However, this equality does not hold for general $f(\cdot)$, so finding the stationary point of $\mathcal{C}_{\rm OG}({\bf w})$ need not yield a consistent estimate of the separating vector. 

To solve this problem, note that $f(\cdot)$ does not appear explicitly in \eqref{eq:grad}. We can therefore consider a replacement of $f(\cdot)$ by its ``normalized'' variant such that the new score function is $\hat\nu^{-1}\phi(\cdot)$, and the new $\nu$ is equal to one. Then, we introduce a modified gradient \eqref{eq:grad} as
\begin{equation}\label{eq:normalizedgrad}
    \nabla={\bf a}-\hat\nu^{-1}\hat{\rm E}\left[\phi\left(\frac{\hat{s}}{\hat\sigma}\right)\frac{\bf x}{\hat\sigma}\right].
\end{equation}
After this modification, the ${\bf w}$ such that $\nabla={\bf 0}$ is a consistent estimate of the true separating vector.

Now, we investigate the second-order derivatives of \eqref{eq:contrastOG}, that is, the derivatives of \eqref{eq:normalizedgrad} in the desired optimum point when $N\rightarrow +\infty$. The result is summarized by the following Proposition.

\begin{theorem}
Let ${\bf z}$ be distributed according to $\mathcal{CN}({\bf 0},{\bf C}_{\bf z})$. Let $f(\cdot)$ be a normalized model pdf so that $\phi(\cdot)\leftarrow\nu^{-1}\phi(\cdot)$, ${\bf w}$ be the true separating vector such that $s={\bf w}^H{\bf x}$, and $N\rightarrow +\infty$. 
Then, the Hessian matrices of \eqref{eq:contrastOG} defined as ${\bf H}_1=\frac{\partial^2\mathcal{C}_{\rm OG}}{\partial{\bf w}^T\partial{\bf w}}$ and
${\bf H}_2=\frac{\partial^2\mathcal{C}_{\rm OG}}{\partial{\bf w}^H\partial{\bf w}}$ are equal to
\begin{align}
    {\bf H}_1&=(c_3{\bf a}{\bf a}^T)^*,\label{eq:H1}\\
    {\bf H}_2&=(c_1\C+c_2{\bf a}{\bf a}^H)^T\label{eq:H2},
\end{align}
where
\begin{align}
    c_1 &= \frac{1}{\sigma^2}\left(\frac{\nu-\rho}{\nu}\right),\label{eq:c1}\\
    c_2 &= -{\sigma^2}c_1-c_3,\label{eq:c2}\\
    c_3 &= \frac{1}{2\nu}(\xi-\eta-\nu)\label{eq:c3},
\end{align}
and
\begin{align}
    \rho &= {\rm E}\left[\frac{\partial\phi(\frac{s}{\sigma})}{\partial s^*} \right],\label{eq:rho}\\
    \xi &= {\rm E}\left[\frac{\partial\phi(\frac{s}{\sigma})}{\partial s^*}\frac{|s|^2}{\sigma^2} \right],\label{eq:xi}\\
    \eta &= {\rm E}\left[\frac{\partial\phi(\frac{s}{\sigma})}{\partial s}\frac{s^2}{\sigma^2} \right]\label{eq:eta}.
\end{align}
\begin{proof}
See Appendix~A.
\end{proof}
\end{theorem}

The proposed one-unit algorithm iterates in the direction inspired by the Newton-Raphson update \cite{li2008}
\begin{equation}\label{eq:newtonraphsonupdate}
    {\bf w}_{\rm new} = {\bf w} - \hat{\bf H}^{-1}(\nabla-\hat{\bf H}_1^*\hat{\bf H}_2^{-1}{\nabla}^*),
\end{equation}
where $\hat{\bf H}=\hat{\bf H}_2^*-\hat{\bf H}_1^*\hat{\bf H}_2^{-1}\hat{\bf H}_1$,  $\nabla$ is given by \eqref{eq:normalizedgrad}, and $\hat{\bf H}_1$ and $\hat{\bf H}_2$ are computed using the expressions \eqref{eq:H1} and \eqref{eq:H2}, respectively, where \eqref{eq:rho}-\eqref{eq:eta} are replaced by their sample-based estimates. That means that the algorithm is not exactly the Newton-Raphson one, because the Hessian matrix is replaced by its analytic expression as if the current ${\bf w}$ was the true separating vector.

In Appendix~B, it is shown that
\begin{equation}\label{eq:H}
    \hat{\bf H}=\left(\frac{\hat\nu-\hat\rho}{\hat\nu}\right)^*\left(\frac{\widehat{\bf C}}{\hat\sigma^2}-{\bf a}{\bf a}^H\right),
\end{equation}
and $\hat{\bf H}_1^*\hat{\bf H}_2^{-1}{\nabla}^*={\bf 0}$, so \eqref{eq:newtonraphsonupdate} is simplified to
\begin{equation}\label{eq:newtonraphsonupdate2}
    {\bf w}_{\rm new} = {\bf w} - \hat{\bf H}^{-1}\nabla.
\end{equation}
However, the reader can notice that $\hat{\bf H}$ is rank deficient, so $\hat{\bf H}^{-1}$ actually does not exist. Indeed, for any value of ${\bf w}$ (and ${\bf a}$ linked through the OGC), the observed signals are equal to ${\bf x}={\bf a}\hat{s}+\hat{\bf y}$ where $\hat{\bf y}={\bf Q}\hat{\bf z}$. The OGC guarantees that $\hat{\rm E}[\hat s \hat{\bf y}]=0$, therefore, $\widehat{\bf C}=\hat\sigma^2{\bf a}{\bf a}^H + \widehat{\bf C}_{\bf y}$,
where $\widehat{\bf C}_{\bf y}=\hat{\rm E}[\hat{\bf y}\hat{\bf y}^H]$. So finally $\hat{\bf H}\propto\widehat{\bf C}_{\bf y}$, whose rank is $d-1$. 
This rank deficiency is caused by the scaling ambiguity of ${\bf w}$: There is a free  scalar parameter with respect to which the contrast function is invariant. Fortunately, it appears that $\nabla$ belongs to the column-space of $\hat{\bf H}$. After some algebra, we receive the following Proposition.
\begin{theorem}\label{theorem:limit}
The update (\ref{eq:newtonraphsonupdate2}) can be re-written as
\begin{equation}\label{eq:newtonupdatefinal}
    {\bf w}_{\rm new} = {\bf w} - \left(\frac{\hat\nu}{\hat\nu-\hat\rho}\right)^*\hat\sigma^2\widehat{\bf C}^{-1}\nabla.
\end{equation}
\begin{proof}
See Appendix~B.
\end{proof}
\end{theorem}

Now, we get back to $T\geq 1$ and $K\geq 1$. By inspecting \eqref{eq:contastCSV}, we can see that all terms with different $t$ values are decoupled. The decoupling also holds for the dataset index $k$ up to the first term in  \eqref{eq:contastCSV}. However, since there is no coupling between the arguments of $f(\{\cdot\}_k)$, we  only need to generalize the definition \eqref{eq:score} to
\begin{equation}
    \phi_{k}\bigl(\bigl\{s^{k,t}\bigr\}_k\bigr)=-\frac{\partial}{\partial {s}_k^*} \log f\bigl(\bigl\{s^{k,t}\bigr\}_k\bigr), 
\end{equation}
and, then, write all the other model parameters and signals' statistics with the superscript ${k,t}$. The gradient of \eqref{eq:contrastOG} and the counterpart of the second-order derivative matrix \eqref{eq:H} are equal to
\begin{align}
    \nabla^k&=\left<{\bf a}^{k,t}-\frac{1}{\hat\nu^{k,t}}\hat{\rm E}\left[\phi_k\left(\left\{\frac{\hat{s}^{k,t}}{\hat\sigma^{k,t}}\right\}_k\right)\frac{{\bf x}^{k,t}}{\hat\sigma^{k,t}}\right]\right>_t,\label{eq:fullgradient}\\
    \hat{\bf H}^k&=\left<\left(\frac{\hat\nu^{k,t}-\hat\rho^{k,t}}{\hat\nu^{k,t}}\right)^*\left(\frac{\widehat{\bf C}^{k,t}}{(\hat\sigma^{k,t})^2}-{\bf a}^{k,t}({\bf a}^{k,t})^H\right)\right>_t.\label{eq:fullhessian}
\end{align}
Similar to \eqref{eq:H}, the scaling ambiguity causes the rank of \eqref{eq:fullhessian} to be exactly equal to $d-1$. However, we can follow the same approach as the one used in Proposition~\ref{theorem:limit} to justify that the update for $T\geq 1$ and $K\geq 1$ is
\begin{equation}\label{eq:updatefinal}
    {\bf w}_{\rm new}^k = {\bf w}^k - \left<\left(\frac{\hat\nu^{k,t}-\hat\rho^{k,t}}{\hat\nu^{k,t}}\right)^*\frac{\widehat{\bf C}^{k,t}}{(\hat\sigma^{k,t})^2}\right>_t^{-1}\nabla^k.
\end{equation}

Given the initial value of ${\bf w}^k$, for all $k=1,\dots,K$, the proposed algorithm  proceeds by computing \eqref{eq:orthogonalconstraint}, $\hat s^{k,t}=({\bf w}^k)^H{\bf x}^{k,t}$, $\hat\sigma^{k,t}$ by \eqref{eq:sigma}, $\hat\nu^{k,t}$ and $\hat\rho^{k,t}$ according to \eqref{eq:nu} and \eqref{eq:rho}, respectively, and updates the separating vectors through \eqref{eq:fullgradient} and \eqref{eq:updatefinal}. The separating vectors can be normalized so that, for example, the scale of the SOI over all blocks equals one. The updates are repeated until the stopping rule from \cite{hyvarinen1999} is satisfied for all $k=1,\dots,K$. The algorithm is referred to as one-unit FastDIVA.

\subsection{Relationship to one-unit FastICA/FastIVA}
One-unit FastICA is designed for BSE for the case $K=1$ and $T=1$ (the indices $k$ and $t$ can be omitted here). When the input signals have been pre-whitened so that $\widehat{\bf C}={\bf I}_d$ \cite{hyvarinen2001}, the one-unit FastICA update rule is 
\begin{equation}\label{eq:oneunitfasticareal}
  {\bf w}_{\rm new} = \hat{\rm E}[\phi(\hat{s}){\bf x}] - \rho{\bf w} 
\end{equation}
for the real-valued case \cite{hyvarinen1999}, and
\begin{equation}\label{eq:oneunitfasticacomplex}
  {\bf w}_{\rm new} = \hat{\rm E}[{\bf x}g(|\hat{s}|^2)] - \hat{\rm E}[g(|\hat{s}|^2) +|\hat{s}|^2g'(|\hat{s}|^2)]{\bf w}
\end{equation}
for the complex-valued case \cite{bingham2000}, where $g(\cdot)$ is the derivative of the contrast function, which is a real-valued smooth even function of $|\hat{s}|^2$. After each update, ${\bf w}$ is normalized, which is equivalent to $\hat\sigma=1$ since $\widehat{\bf C}={\bf I}$. 

We can compare \eqref{eq:updatefinal} in a similar setting when $\hat\sigma=1$ and $\widehat{\bf C}={\bf I}_d$. The OGC \eqref{eq:orthogonalconstraint} is then translated to ${\bf a}={\bf w}$, and \eqref{eq:updatefinal} is simplified to
\begin{equation}\label{eq:oneunitfasticalike}
  {\bf w}_{\rm new} = {\bf w} - \left(\frac{\hat\nu}{\hat\nu-\hat\rho}\right)^*({\bf w}-\hat\nu^{-1}\hat{\rm E}\left[\phi\left(\hat{s}\right){\bf x}\right]),  
\end{equation}
Since the scale of ${\bf w}_{\rm new}$ can be arbitrary (the vector can be normalized afterwards), the right-hand side of \eqref{eq:oneunitfasticalike} can be multiplied by the scalar factor $(\hat\nu-\hat\rho)^*$, which, after a few simplifications, results in
\begin{equation}\label{eq:oneunitfasticalike2}
  {\bf w}_{\rm new} = \hat{\rm E}[\phi(\hat{s}){\bf x}] - \rho^*{\bf w}. 
\end{equation}
It is worth noting here that $\rho$ should be real-valued, provided that the model density $f(\cdot)$ is a real-valued function. Once $f(\cdot)=f(\cdot)^*$, it holds that $\rho=\rho^*$ \cite{kreutzdelgado2009}.
By comparing \eqref{eq:oneunitfasticalike2} with \eqref{eq:oneunitfasticareal}, we can see that the update rules of one-unit FastDIVA and one-unit FastICA are the same in the real-valued case. 

The complex-valued FastICA was derived in a different way, assuming a constrained class of contrast functions suitable for circular sources. The update rule \eqref{eq:oneunitfasticacomplex} is different from  \eqref{eq:oneunitfasticalike2}. The latter is actually simpler and valid for circular as well as non-circular SOI (and a circular background). 

For $T=1$ and $K\geq 1$, similar conclusions hold when comparing the update rules of FastIVA derived in \cite{lee2007fast} (Equation~58 in \cite{lee2007fast}), which are similar to \eqref{eq:oneunitfasticacomplex}, while \eqref{eq:updatefinal} is simplified (when $T=1$, $\widehat{\bf C}^k={\bf I}_d$ and $\sigma^k=1$) to
\begin{equation}\label{eq:updateIVE}
    {\bf w}_{\rm new}^k = \hat{\rm E}[\phi_k(\{\hat{s}^k\}_k){\bf x}^k]) - (\rho^k)^*{\bf w}^k.
\end{equation}

To conclude, one-unit FastDIVA is an extension of FastICA and FastIVA for $T>1$ under the CSV model, in the real-valued case, and an extension and simplification involving  non-circular SOI, in the complex-valued case.

\subsection{Separation of several signals}\label{section:BSS}
We now focus on the BSS problem when $1 \leq r \leq d$ independent signals should be separated from each other and from the remainder of the signal (i.e., the other components and the noise). Following the idea of \cite{delfosse1995,hyvarinen1999}, we propose to run $r$ one-unit algorithms successively or in parallel while preventing them from extracting the same sources. To this end, the orthogonality constraint is imposed \cite{cardoso1994}.

Throughout this Subsection, we will omit the dataset index $k$ as the proposed approaches operate independently in each dataset.

\subsubsection{Symmetric approach}
The approach presented here is suitable for dynamic mixtures \eqref{eq:mixingmodel} satisfying condition (C1) as defined in Section~\ref{section:separable}. The deflation and symmetric approaches can then be used as they were designed for the static case $T=1$ \cite{hyvarinen1999}.
Let us recall the symmetric approach here (the deflation approach can be derived similarly \cite{hyvarinen1999}). 

Consider $r$ separating vectors ${\bf w}_1,\dots,{\bf w}_r$ each being updated through \eqref{eq:updatefinal}. Since the output signals, denoted as $\hat s_1^t={\bf w}_1^H{\bf x}^t,\dots,\hat s_r^t={\bf w}_r^H{\bf x}^t$, should be independent, it is reasonable that their mutual correlations estimated over all available samples (and blocks) should be constrained to equal zero. Specifically, the condition is that
\begin{equation}\label{eq:OGcondition1}
\bigl<\hat{\rm E}[\hat s_i^t (\hat s_j^t)^*]\bigr>_t=\bigl<{\bf w}_i^H\widehat{\bf C}^t{\bf w}_j\bigr>_t={\bf w}_i^H{\bf R}{\bf w}_j=\delta_{ij}, 
\end{equation}
where ${\bf R}=\bigl<\widehat{\bf C}^t\bigr>_t$ and $\delta_{ij}$ denotes the Kronecker symbol, and $i,j=1,\dots,r$. Let ${\bf W}^+=[{\bf w}_1,\dots,{\bf w}_r]$ involve the separating vectors after they were updated through \eqref{eq:updatefinal}, which do not satisfy \eqref{eq:OGcondition1}, in general. The symmetric approach therefore proceeds by
\begin{equation}\label{eq:symmetricorthogonalization}
{\bf W}_{\rm new} = {\bf W}^+\left(({\bf W}^+)^H{\bf R}{\bf W}^+\right)^{-\frac{1}{2}}.
\end{equation}
Since ${\bf W}_{\rm new}^H{\bf R}{\bf W}_{\rm new}={\bf I}_r$, the columns of ${\bf W}_{\rm new}$ satisfy \eqref{eq:OGcondition1} and, therefore, can be used as the orthogonalized counterparts of ${\bf W}^+$. 

Symmetric FastDIVA, as the proposed method to separate $r$ independent signals is called,  alternates between the updates of the separating vectors according to \eqref{eq:updateIVE} and their subsequent orthogonalizations \eqref{eq:symmetricorthogonalization}, until convergence.

\subsubsection{Block-Deflation approach} 
This approach is tailored to mixtures \eqref{eq:mixingmodel} satisfying condition (C2) as defined in Section~\ref{section:separable}. It imposes a stronger condition on the extracted signals by making them orthogonal separately in each block. Specifically, it is expected that 
\begin{equation}\label{eq:OGcondition2}
\hat{\rm E}[\hat s_i^t (\hat s_j^t)^*]=\delta_{ij}(\hat\sigma^{k,t}_i)^2
\end{equation}
for every $t=1,\dots,T$ and $i,j=1,\dots,r$.

To this end, we propose an extended, so-called, {\em block deflation} scheme, which proceeds as follows. The first signal is extracted from the original data by one-unit FastDIVA. The extracted signal is then subtracted from the original input signals (on each block) using least-squares projections. Then, one-unit FastDIVA is applied to the new data and extracts the second signal, whose orthogonality is ensured due to the projection properties. This process is repeated recursively until $r$ signals are extracted. 

Let ${\bf x}_i^t$ denote the input signals on the $t$th block at the $i$th stage of the block-deflation scheme, and let ${\bf w}_i$ be the separating vector obtained after one iteration by one-unit FastDIVA applied to ${\bf x}_i^t$. For $i=1$, ${\bf x}_i^t={\bf x}^t$ (the original input data). The new data ${\bf x}_{i+1}^t$ are obtained by the least-squares subtraction of $\hat s_i^t={\bf w}_i^H{\bf x}_i^t$ from ${\bf x}_i^t$. Owing to the OGC \eqref{eq:orthogonalconstraint} imposed between the mixing and separating vectors of the extracted source, the new data is obtained through
\begin{equation}
    {\bf x}_{i+1}^t = \boldsymbol{\Pi}_{i}^t {\bf x}^t_i, 
\end{equation}
where $\boldsymbol{\Pi}_{i}^t = {\bf E}_i\left({\bf I}_{d-i+1}-\dot{\bf a}_i^t{\bf w}_i^H\right)$; $\dot{\bf a}_i^t$ is the estimated mixing vector on the $t$th block corresponding to the $i$th extracted signal with respect to data ${\bf x}_i^t$. ${\bf E}_i$ is a suitable $(d-i)\times (d-i+1)$ matrix having the full row-rank; it reduces the dimension of ${\bf x}_{i+1}^t$ as compared to ${\bf x}^t_i$ by one (so that the new data is not rank deficient); the dimension of ${\bf x}_i^t$ is $d-i+1$. 

The estimated vectors ${\bf w}_i$ and $\dot{\bf a}_i^t$ operate on the data ${\bf x}^t_i$. In order to derive their counterparts operating on the original data ${\bf x}^t$, let us introduce the following definitions:
\begin{align}
    {\bf P}_1^t&={\bf I}_d,\label{eq:bdstep1}\\
    {\bf P}_i^t &= \boldsymbol{\Pi}_{i-1}^t\boldsymbol{\Pi}_{i-2}^t\dots \boldsymbol{\Pi}_{1}^t,\quad i>1\label{eq:bdstep2}\\
    \widehat{\bf C}_i^t &= \hat{\rm E}[{\bf x}_i^t({\bf x}_i^t)^H],\\
    {\bf w}_i^t &= ({\bf P}_i^t)^H{\bf w}_i,\label{eq:bdstep3}\\
    {\bf a}_i^t &= \frac{\widehat{\bf C}^t{\bf w}_i^t}{({\bf w}_i^t)^H\widehat{\bf C}^t{\bf w}_i^t}.\label{eq:OGCinoriginaldomain}
\end{align}
It is then straightforward to verify that, for $i=1,\dots,r$,
\begin{align}
    {\bf x}_{i}^t &= {\bf P}_i^t {\bf x}^t,\\
    \hat s_i^t &={\bf w}_i^H{\bf x}_i^t = ({\bf w}_i^t)^H{\bf x}^t,\\
    \widehat{\bf C}_i^t &= {\bf P}_i^t\widehat{\bf C}^t({\bf P}_i^t)^H,\\
    \dot{\bf a}_i^t &= {\bf P}_i^t {\bf a}_i^t.
\end{align}
Note that ${\bf w}_i$ and $\dot{\bf a}_i^t$ operate on ${\bf x}_i^t$ and, since both have been estimated by one-unit FastDIVA, they are coupled through the OGC, i.e., $\dot{\bf a}_i^t=\frac{\widehat{\bf C}_i^t{\bf w}_i}{({\bf w}_i)^H\widehat{\bf C}_i^t{\bf w}_i}$. In addition, ${\bf w}_i$ is independent of $t$ due to the CSV model assumed by one-unit FastDIVA. 

The counterpart of ${\bf w}_i$ and $\dot{\bf a}_i^t$ operating on ${\bf x}^t$ is ${\bf w}_i^t$ and ${\bf a}_i^t$, respectively. Interestingly, unless $i=1$ holds, ${\bf w}_i^t$ is, in general, no longer independent of $t$.

\section{Performance Analysis}\label{section:analysis}
The goal here is to analyze the accuracy of one-unit FastDIVA considering the BSE problem under the CSV mixture model. The accuracy is studied by analyzing the mean residual presence of the $j$th original signal in the extracted signal $j=1,\dots,d$, which is characterized by the mean  interference-to-signal ratio (ISR) achieved by the algorithm. 

To this end, we compute the asymptotic variance of the estimated separating vector that is obtained by the algorithm as the optimum point of the contrast function \eqref{eq:contrastOG}; it is assumed that $N\rightarrow \infty$, which means, for a fixed value of $T$, that also $N_b\rightarrow \infty$. Using the equivariance property of the BSE problem, proven in \cite{kautsky2020CRLB}, we consider the special case as if the true mixing and separating vectors were ${\bf a}^{k,t}={\bf w}^k=[1;{\bf 0}]$, $k=1,\dots,K$ (in Section~\ref{section:simulations}, this analysis is verified for general mixing and separating vectors). Then, by \eqref{eq:mixingmatrixIVE} it follows that ${\bf x}^{k,t}=[s^{k,t};-{\bf z}^{k,t}]$.

Let $\hat{\bf w}^k$, $\hat{s}^{k,t}=(\hat{\bf w}^k)^H{\bf x}^{k,t}$, $\hat{\sigma}_s^{k,t}$ and $\hat{\bf z}^{k,t}$ denote, respectively, the estimates of ${\bf w}^k$, $s^{k,t}$, of the sample-based variance estimate of $\hat{s}^{k,t}$, and of the background signals. The following notation will be used:
\begin{eqnarray}
\hat\nu_s^{k,t}&=& \hat{\rm E}\left[\phi_k\left(\left\{\frac{\hat{s}^{k,t}}{\hat\sigma^{k,t}_s}\right\}_k\right)\frac{\hat{s}^{k,t}}{\hat\sigma_s^{k,t}}\right],\\
\hat\rho_s^{k,t}&=& \hat{\rm E}\left[\frac{\partial \phi_k}{\partial s_{k}^*}\left(\left\{ \frac{{\hat{s}}^{k,t}}{\hat\sigma_s^{k,t}}\right\}_k\right)\right].
\end{eqnarray} 
Next, we introduce the random variables derived from the samples of $s^{k,t}$ and ${\bf z}^{k,t}$
\begin{eqnarray}
\hat{\boldsymbol \chi}^{k,t}&=& \hat{\rm E}\left[{\bf z}^{k,t}(s^{k,t})^H\right]\\
\hat{{\boldsymbol \zeta}}^{k,t} &=& \hat{\rm E}\left[\phi_k\left(\left\{\frac{{ s}^{k,t}}{\sigma^{k,t}}\right\}_k\right)\frac{{\bf z}^{k,t}}{\sigma^{k,t}}\right]\\
\hat{\boldsymbol \zeta}_s^{k,t} &=& \hat{\rm E}\left[\phi_k\left(\left\{\frac{\hat{{ s}}^{k,t}}{\hat{\sigma}^{k,t}_s}\right\}_k\right)\frac{{\bf z}^{k,t}}{\hat\sigma^{k,t}_s}\right].
\end{eqnarray}
Let the structure of $\hat{\bf w}^k$ be
\begin{equation}
    \hat{\bf w}^k=[1+p^k,{\bf q}^k]^T
\end{equation}
where $p^k$ and ${\bf q}^k$ are random variables of the stochastic order $O_p(N_b^{-1})$ and $O_p(N_b^{-1/2})$, respectively; $O_p(\cdot)$ represents the stochastic order symbol; see Appendix~C in \cite{Porat2008}.
The goal now is to express $p^k$ and ${\bf q}^k$ as functions of $s^{k,t}$ and ${\bf z}^{k,t}$ and to compute their asymptotic variances. Finally, only the asymptotic covariance of ${\bf q}^k$ will be needed.

Note that $\hat{\boldsymbol \chi}^{k,t}$ and $\hat{\boldsymbol \zeta}^{k,t}$ have the same stochastic order below.
We can write
\begin{equation}
\widehat{\bf C}^{k,t} = \hat{\rm E}[{\bf x}^{k,t}({\bf x}^{k,t})^H]= \left[\begin{array}{cc} (\sigma^{k,t})^2+c^{k,t} & -(\hat{\boldsymbol \chi}^{k,t})^H\\ -\hat{\boldsymbol \chi}^{k,t} & {\bf C}_{\bf z}^{k,t}+ \boldsymbol{\Xi}^{k,t}\end{array}\right],
\end{equation} where
\begin{align}
c^{k,t} =&\ \hat{\rm E}[s^{k,t}(s^{k,t})^H]-(\sigma^{k,t})^2,\\
\boldsymbol{\Xi}^{k,t} =&\ \hat{\rm E}[{\bf z}^{k,t}({\bf z}^{k,t})^H]-{\bf C}_{\bf z}^{k,t}.
\end{align}
Define the difference between the sample-based variances as
\begin{multline}
    b^{k,t} = \hat{\sigma}^{k,t} - \hat{\sigma}_s^{k,t} = \hat{\rm E}[s^{k,t}(s^{k,t})^H] - \hat{\rm E}[\hat{s}^{k,t}(\hat{s}^{k,t})^H] 
    \\ = 
    \hat{\rm E}[({\bf q}^k)^H {\bf z}^{k,t}({\bf z}^{k,t})^H{\bf q}^k].
\end{multline}
The stochastic order of $c^{k,t}$ and $\boldsymbol{\Xi}^{k,t}$ is the same as that of $\hat{\boldsymbol \chi}^{k,t}$, i.e., $O_p(N_b^{-1/2})$, while $b^{k,t}$ is of order $O_p(N_b^{-1})$. It holds that
\begin{equation}
\frac{\hat{s}^{k,t}}{\hat{\sigma}_s^{k,t}}= \frac{(\hat{\bf w}^k)^H{\bf x}^{k,t}}{ \sigma^{k,t}+b^{k,t}+c^{k,t}}= \frac{s^{k,t}}{\sigma^{k,t}}
-\frac{({\bf q}^k)^H{\bf z}^{k,t}}{\sigma^{k,t}}
+ o_p(N_b^{-1/2}).
\end{equation}
Assuming the smoothness of $\phi$ and using the first-order Taylor series expansion, we get
\begin{multline}
\phi_k\left(\left\{\frac{{\hat{ s}}^{k,t}}{{\hat{\sigma}}^{k,t}_s}\right\}_{k}\right)
=  \phi_k\left(\left\{\frac{{ s}^{k,t}}{{\sigma}^{k,t}}\right\}_{k}\right) \\
 - \frac{1}{\sigma^{k,t}}\sum_{k=1}^K({\bf q}^k)^H{\bf z}^{k,t}\frac{\partial \phi_k}{\partial s_k}\left(\left\{\frac{{ s}^{k,t}}{{\sigma}^{k,t}}\right\}_{k}\right) 
 \\ - 
 \frac{1}{\sigma^{k,t}}\sum_{k=1}^K({\bf z}^{k,t})^H{\bf q}^k
 \frac{\partial \phi_k^*}{\partial s_k}\left(\left\{\frac{{ s}^{k,t}}{{\sigma}^{k,t}}\right\}_{k}\right) + o_p(N_b^{-1/2}).
\end{multline}
From the uncorrelatedness of datasets, and assuming the circularity of ${\bf z}^{k,t}$, i.e ${\rm E}\left[{\bf z}^{k,t}({\bf z}^{k,t})^T\right]={\bf 0}$, we can write
\begin{multline}\label{an:zeta}
\hat{\boldsymbol \zeta}_s^{k,t} = \hat{\rm E}\left[\phi_k\left(\left\{\frac{\hat{{ s}}^{k,t}}{\hat{\sigma}^{k,t}}\right\}_k\right)\frac{{\bf z}^{k,t}}{\hat{\sigma}^{k,t}}\right] \\ =
\hat{\rm E}\left[\phi_k\left(\left\{\frac{{{ s}}^{k,t}}{\sigma^{k,t}}\right\}_k\right)\frac{{\bf z}^{k,t}}{\sigma^{k,t}} \right] \\ - \frac{1}{(\sigma^{k,t})^2}\hat{\rm E}\left[\frac{\partial \phi_k}{\partial s_k}\left(\left\{\frac{{{s}}^{k,t}}{\sigma^{k,t}}\right\}_k\right){\bf z}^{k,t}({\bf z}^{k,t})^T({\bf q}^k)^* \right] 
\\ - \frac{1}{(\sigma^{k,t})^2}\hat{\rm E}\left[\frac{\partial \phi_k}{\partial s_k^*}\left(\left\{\frac{{{s}}^{k,t}}{\sigma^{k,t}}\right\}_k\right){\bf z}^{k,t}({\bf z}^{k,t})^H{\bf q}^k\Big]\right] 
+ 
o_p(N_b^{-1/2}) 
\\
 = \hat{\boldsymbol \zeta}^{k,t}-\frac{\hat{\rho}^{k,t} {\bf C}_{\bf z}^{k,t}{\bf q}^k}{(\sigma^{k,t})^2} 
+o_p(N_b^{-1/2}).
\end{multline} 
Next,
\begin{align}
\widehat{\bf C}^{k,t} {\bf w}^k =&\ \left[\begin{array}{c} (\sigma^{k,t})^2+c^{k,t}\\ -\hat{\boldsymbol \chi}^{k,t}+{\bf C}_{\bf z}^{k,t}{\bf q}^k\end{array}\right]+o_p(N_b^{-1/2})\\
({\bf w}^k)^H \widehat{\bf C}^{k,t} {\bf w}^k =&\ (\sigma^{k,t})^2+c^{k,t}+o_p(N_b^{-1/2}).\label{an:wCw}
\end{align}

The  mixing vector estimated by \eqref{eq:orthogonalconstraint} and the gradient \eqref{eq:fullgradient} can be expressed, using \eqref{an:zeta}-\eqref{an:wCw}, respectively, as
\begin{align}
\hat{\bf a}^{k,t} =&\ \frac{1}{(\sigma^{k,t})^2}\left[\begin{array}{c} (\sigma^{k,t})^2\\ -\hat{\boldsymbol \chi}^{k,t}+{\bf C}_{\bf z}^{k,t}{\bf q}^k\end{array}\right]+o_p(N_b^{-1/2}),\\
\widehat{\nabla}^k =&\ \left< {\bf a}^{k,t}-\left[\begin{array}{c} 1\\ -\hat{\boldsymbol \zeta}_s^{k,t}/\hat{\nu}_s^{k,t}\end{array}\right] \right>_t \\
=&\ \left<\left[\begin{array}{c} 0 \nonumber \\ \frac{-\hat{\boldsymbol \chi}^{k,t} + {\bf C}_{\bf z}^{k,t}{\bf q}^k}{(\sigma^{k,t})^2} +\hat{\boldsymbol \zeta}_s^{k,t}/ \hat{\nu}_s^{k,t}\end{array}\right]\right>_t
+o_p(N^{-1/2}).
\end{align}
The stationary point of the algorithm is now sought as the solution of $\widehat{\nabla}^k={\bf 0}$. This gives us
\begin{equation}\label{an:q}
    {\bf q}^k = {\bf R}^{k,t}
    \left<\frac{{\hat{\nu}^{k,t}\hat{\boldsymbol \chi}^{k,t}-\hat{\boldsymbol \zeta}^{k,t}(\sigma^{k,t})^2}}{(\sigma^{k,t})^2 \hat{\nu}^{k,t}}\right>_t 
    +o_p(N^{-1/2}),
\end{equation}
where
\begin{equation}\label{an:R}
    {\bf R}^{k,t}=\left<{\bf C}_{\bf z}^{k,t}\frac{{{\nu}^{k,t}-{\rho}^{k,t}}}{(\sigma^{k,t})^2 {\nu}^{k,t}}\right>_t^{-1}.
\end{equation}

Computation of the asymptotic covariance of ${\bf q}^k$ remains to be done.
Straightforward computations give
\begin{eqnarray}
{\rm E}[\hat{\boldsymbol \chi}^{k,t}(\hat{\boldsymbol \chi}^{k,t})^H]&=& \frac{1}{N_b}(\sigma^{k,t})^2{\bf C}_{\bf z}^{k,t}\\
{\rm E}[\hat{\boldsymbol \zeta}^{k,t}(\hat{\boldsymbol \zeta}^{k,t})^H]&=&\frac{1}{N_b}\frac{{\varphi}^{k,t}}{(\sigma^{k,t})^2}{\bf C}_{\bf z}^{k,t}\\
{\rm E}[\hat{\boldsymbol \chi}^{k,t}(\hat{\boldsymbol \zeta}^{k,t})^H]&=&\frac{{\nu}^{k,t}}{N_b}{\bf C}_{\bf z}^{k,t}, 
\end{eqnarray}
where we have introduced one more statistic related to the SOI
\begin{equation}
   {\varphi}^{k,t}= {\rm E}\left[\left|\phi_k\left(\left\{\frac{{ s}^{k,t}}{{\sigma}_s^{k,t}}\right\}_k\right)\right|^2\right].
\end{equation}
Using these expressions and \eqref{an:q}, the asymptotic covariance of ${\bf q}^k$ is given by
\begin{multline}\label{an:cov}
{\tt cov}[{\bf q}^k] = {{\tt cov}\left[{\bf R}^{k,t}\left<\frac{\hat{\nu}^{k,t}\hat{\boldsymbol \chi}^{k,t}-\hat{\boldsymbol \zeta}^{k,t}(\sigma^{k,t})^2}{(\sigma^{k,t})^2 \hat{\nu}^{k,t}}\right>_t\right]}\\
 =\frac{1}{N}{\bf R}^{k,t}{\left<{\bf C}_{\bf z}^{k,t}\frac{{\varphi}^{k,t}-|{\nu}^{k,t}|^2}{(\sigma^{k,t})^2 |{\nu}^{k,t}|^2}\right>_t}({\bf R}^{k,t})^* + o(N^{-1}).
\end{multline}
The theoretical ISR reads 
\begin{multline}\label{eq:ISRdef}
    {\tt ISR}^{k}  =
    \frac{\sum_{t=1}^{T}{\rm E}\left[ \left| (\widehat{\bf w}^k)^H {\bf y}^{k,t} \right|^2\right]}{\sum_{t=1}^{T}{\rm E}\left[ \left| (\widehat{\bf w}^k)^H {\bf a}^{k,t} s^{k,t} \right|^2\right]} = \\
     \frac{\sum_{t=1}^{T}\tr\left[  {\bf C}_{\bf z}^{k,t} {\tt cov}[{\bf q}^k] \right]}{\sum_{t=1}^{T}(\sigma^{k,t})^2 
     } = \frac{\tr\left[  \left<{\bf C}_{\bf z}^{k,t}\right>_t {\tt cov}[{\bf q}^k] \right]}
     {\left<(\sigma^{k,t})^2\right>_t}.
\end{multline} 
Hence, using \eqref{an:R} and \eqref{an:cov}, the asymptotic mean ISR achieved by the algorithm is
\begin{multline}\label{eq:finalISR}
     {\rm E}\left[{\tt ISR}^{k}\right] \approx 
     \frac{1}{N}\tr \Bigg[
     \frac{\left<{\bf C}_{\bf z}^{k,t}\right>_t}{\left<(\sigma^{k,t})^2\right>_t}
     \left<{\bf C}_{\bf z}^{k,t}\frac{{{\nu}^{k,t}-{\rho}^{k,t}}}{(\sigma^{k,t})^2 {\nu}^{k,t}}\right>_t^{-1}
     \\
     {\left<{\bf C}_{\bf z}^{k,t}\frac{{\varphi}^{k,t}-|{\nu}^{k,t}|^2}{(\sigma^{k,t})^2 |{\nu}^{k,t}|^2}\right>_t}
     \left(\left<{\bf C}_{\bf z}^{k,t}\frac{{{\nu}^{k,t}-{\rho}^{k,t}}}{(\sigma^{k,t})^2 {\nu}^{k,t}}\right>_t^{-1}\right)^*
     \Bigg].
\end{multline} 

To compare this result with previous analyses, consider $T=K=1$. Then,  \eqref{eq:finalISR} is simplified to 
\begin{equation}
    \mbox{E}\left[{\tt ISR}^{k}\right] \approx \frac{d-1}{N}\frac{{\varphi}^{k}-|{\nu}^{k}|^2}{|{\nu}^{k}-{\rho}^{k}|^2},
\end{equation} 
which coincides with the results given in \cite{laheld1996, hyvarinen1997b, tichavsky2006} (that result is also confirmed in the present paper for the complex-valued case and for $K>1$).

Next, let the model density $f(\cdot)$ correspond to the normalized true pdf of the SOI for all $k$ and $t$; let us, for the moment, denote this normalized true pdf by $p^{k,t}(\cdot)$. Then the equalities $\nu^{k,t} = 1$, $\rho^{k,t} = \kappa^{k,t}$ and $\varphi^{k,t}=\kappa^{k,t}$ hold, where 
\begin{equation}
    \kappa^{k,t} = \mbox{E}\left[\left|\frac{\partial \log p^{k,t}\bigl(\{s_k\}_k\bigr)}{\partial s_k^*}\right|^2\right].
\end{equation}
Formula \eqref{eq:finalISR} now takes on the form
\begin{equation}\label{eq:CRLB}
    \mbox{E}\left[{\tt ISR}^{k}\right] \approx
     \frac{1}{N}\tr \left[
     \frac{\left<{\bf C}_{\bf z}^{k,t}\right>_t}{\left<(\sigma^{k,t})^2\right>_t}
     \left<{\bf C}_{\bf z}^{k,t}\frac{\kappa^{k,t}-1}{(\sigma^{k,t})^2 }\right>_t^{-1} \right].
\end{equation} 
For $K=1$, \eqref{eq:CRLB} coincides with the Cram\'er-Rao Lower Bound  derived in \cite{kautsky2020CRLB} (Eq. 70 in \cite{kautsky2020CRLB}), which points to the asymptotic efficiency of one-unit FastDIVA under the corresponding statistical (and mixing) model when the used nonlinearity corresponds with the true normalized score function of the SOI.

\section{Numerical Validation}\label{section:simulations}
In experiments, we simulate BSE and BSS on mixtures obeying dynamic models discussed in this paper. In BSE, one-unit FastDIVA is compared with recent methods assuming CSV mixing, namely, with the gradient-based BOGIVE$_{\bf w}$ \cite{koldovsky2019icassp} and with a more advanced QuickIVE-2 \cite{koldovsky2019quick}. FastICA/FastIVA are compared in the BSS tasks, both implemented as FastDIVA with a special setting (i.e., when $T=1$ is assumed). All the algorithms use the rational nonlinearity given by \cite{rati2007}
\begin{equation}\label{eq:ratinonl}
    \phi_k(\{s_k\}_k)=\frac{s_k^*}{1+\sum_{k=1}^K|s_k|^2}.
\end{equation}
The number of iterations is restricted to $100$ in QuickIVE-2 and in FastDIVA and to $1,000$ in BOGIVE$_{\bf w}$. The step size in BOGIVE$_{\bf w}$ is set to $0.1$.

The accuracy of separated signals is evaluated, after resolving the unknown order, in terms of ISR, as defined by the first fraction in \eqref{eq:ISRdef} (the expectations are replaced by sample averages).

\subsection{Dynamic Blind Source Extraction}
The simulation here is focused on the BSE problem to verify the efficiency of one-unit FastDIVA, to verify its analysis provided in Section~\ref{section:analysis}, and to evaluate its speed. In a trial, a mixture of dimension $d=6$ is generated such that it obeys CSV with $T=5$ blocks of length $N_b=2,000$, i.e., $N=10^4$. The background signals are circular  Gaussian while the SOI is generated according to the complex-valued Generalized Gaussian distribution \cite{loeschCRB} with the shape parameter $\alpha$, denoted as GG($\alpha$). The variance of SOI is block-dependent, namely,  equal to $|\cos(i/6*\pi)| + 1 - \sqrt{3}/2$ on the $i$th block. The mixing matrices are randomly generated so that the first rows of their inverse matrices are the same in all blocks, that is, \eqref{eq:CSV} is satisfied. 

The experiment is realized in two variants with $K=1$ and $K=2$. In the latter case, the SOIs  are, in both mixtures, rotated by a random unitary matrix (before they are mixed with the background) in order to establish their higher-order dependence. The compared methods are initialized by randomly perturbed true separating vectors, where the elements of the perturbations are $\mathcal{CN}(0,0.1)$.

\begin{figure}
    \centering
    \includegraphics[width=\linewidth]{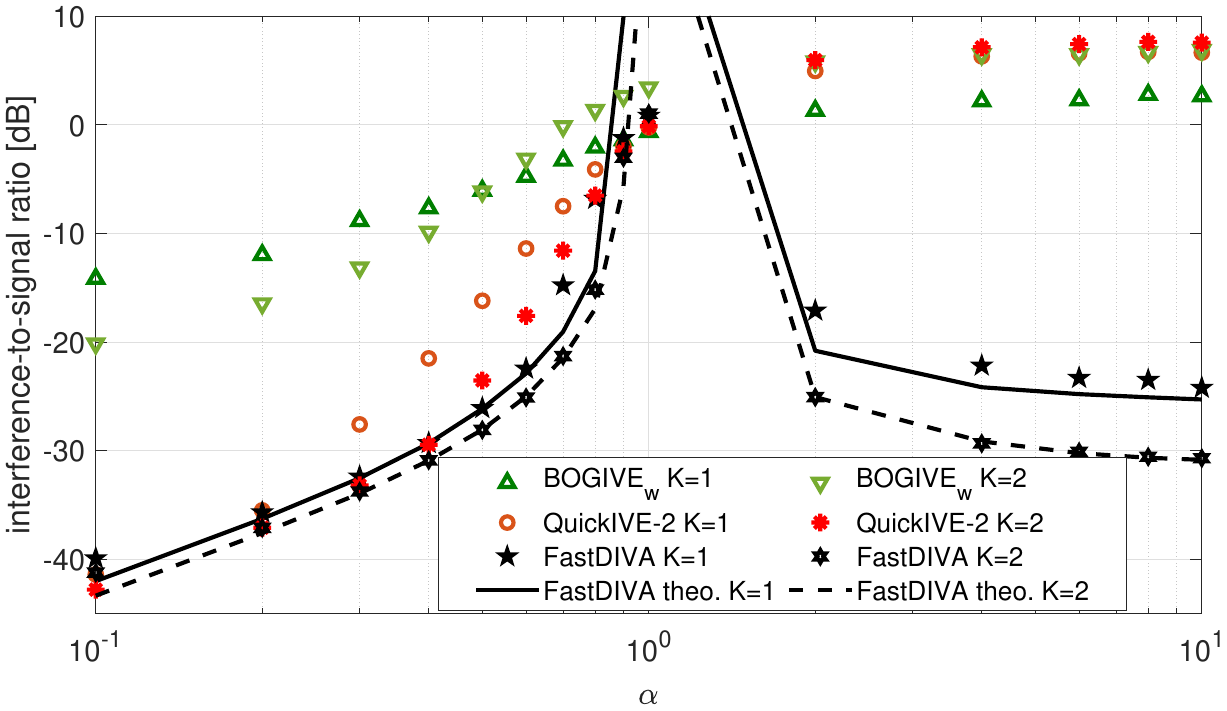}
    \caption{Average ISR over $1000$ as a function of $\alpha$, the shape parameter of the pdf of the SOI; $\alpha=1$ corresponds to Gaussian SOI, which is not identifiable. The pdf super-Gaussian and sub-Gaussian for $\alpha<1$ and $\alpha>1$, respectively. "FastDIVA theo." stands for the analytical prediction \eqref{eq:finalISR}.}
    \label{fig:fig1}
\end{figure}

Fig.~\ref{fig:fig1} shows ISR averaged\footnote{One percent of minimum and maximum values of ISR were discarded in order to eliminate the bias caused by the ambiguity of order (the algorithm might, in a few trials, be attracted by a different extreme of the contrast function corresponding to a  signal different from the SOI.).} over $1,000$ trials as a function of $\alpha\in[0.1,10]$. Note that the SOI is super-Gaussian for $\alpha<1$, Gaussian for $\alpha=1$, and sub-Gaussian for $\alpha>1$. For $\alpha=1$, the SOI is not identifiable. The average ISR established by the methods therefore tends to be close to or above $0$~dB when $\alpha$ is close to one, which means a poor extraction accuracy. 

One-unit FastDIVA yields performance that is in good agreement with the theoretical analysis given by \eqref{eq:finalISR}. BOGIVE$_{\bf w}$ gives poor ISR compared to the other methods, because $1,000$ iterations is generally not sufficient to achieve the optimum point. QuickIVE-2 achieves  results similar to FastDIVA for $\alpha<0.3$ and slightly worse for $\alpha\in[0.3,1]$ (also because of the limited number of iterations). For $\alpha>1$, BOGIVE$_{\bf w}$ and QuickIVE-2 fail to extract the SOI since the algorithms are not stable with respect to the SOI sub-Gaussianity and the nonlinearity \eqref{eq:ratinonl}. Here, FastDIVA inherits the stability of FastICA and works well also for $\alpha>1$.

\begin{figure}
    \centering
    \includegraphics[width=\linewidth]{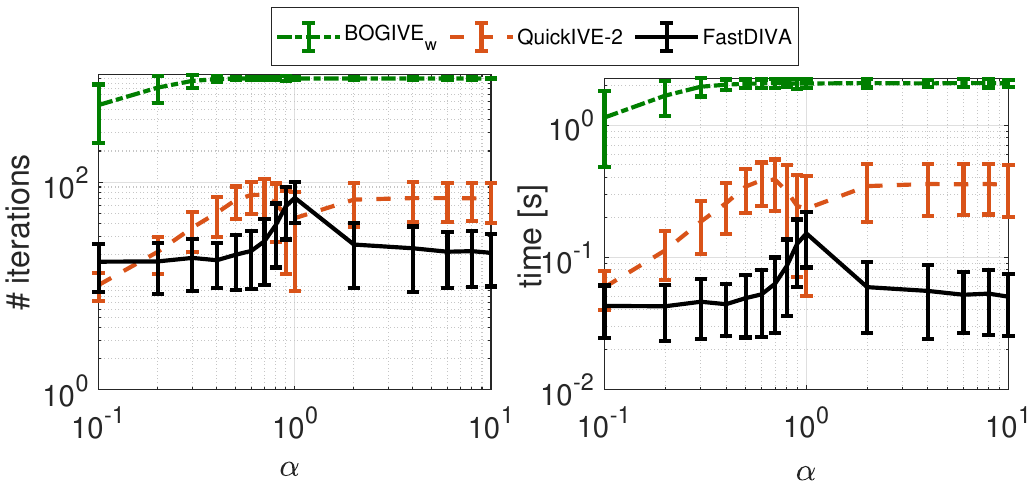}
    \caption{The number of iterations and computational time needed by the compared methods to achieve convergence. Note that the maximum number of iteration is $100$ for FastDIVA and QuickIVE-2 and $1000$ for BOGIVE$_{\bf w}$; simulations were done in Matlab R2020a on a server with Inter Xeon 12-core 2.6 GHz CPU, 64 GB RAM.}
    \label{fig:fig2}
\end{figure}

For $K=2$, all methods achieve improved ISR as compared to the case of $K=1$, which confirms the advantage following from the joint source extraction \cite{kautsky2019icassp}. Fig.~\ref{fig:fig2} shows the computational complexity in terms of the number of iterations and computational time. FastDIVA and QuickIVE-2 show significantly faster convergence as compared to BOGIVE$_{\bf w}$, and  FastDIVA is faster than QuickIVE-2.

\subsection{Dynamic Separation of Several Sources}
Now, we focus on the BSS problem of $1\leq r\leq d$ signals from mixtures of dimension $d$ obeying condition (C1) or (C2), as defined in Section~\ref{section:separable}. In this scenario, $d=5$, $T=5$, $K=1$, $N_b=10^4$, $N=5\cdot 10^4$.

As for (C1), $r$ complex-valued signals are generated according to GG($0.1$) with the same variance profiles as the SOI in the previous experiment. The background is considered in two variants: Gaussian or GG($0.1$). The mixing matrices are randomly generated so that the first $r$ rows of their inverse matrices are the same in all blocks. The mixing model is static for $r=d$.

In the case of (C2), real-valued mixtures of the speech signals from Fig.~\ref{fig:speechsignals} are considered. The mixing matrices are generated as follows. In the beginning, mixing and separating vectors $\dot{\bf a}_i$ and ${\bf w}_i$ of dimension $i$ are generated at random, where $i=1,\dots,d$ such that  ${\bf w}_i^H\dot{\bf a}_i=1$; their values remain fixed during the rest of the simulations. Then, in a trial, their values are perturbed by random vectors of the same size whose elements are taken from $\mathcal{N}(0,\lambda^2)$. The mixing vector $\dot{\bf a}_i$ is perturbed differently on each block, which simulates a random walk of the associated source; $\lambda$ thus plays the role of a variability coefficient of the mixture. The rows of de-mixing matrices are then obtained successively by using \eqref{eq:bdstep1}, \eqref{eq:bdstep2} and \eqref{eq:bdstep3}. The mixing matrices are obtained as the inverse matrices of the de-mixing ones for $i=1,\dots,d$. These steps guarantee that the mixtures obey (C2); for $\lambda=0$, they are static.

\begin{figure}
    \centering
    \includegraphics[width=\linewidth]{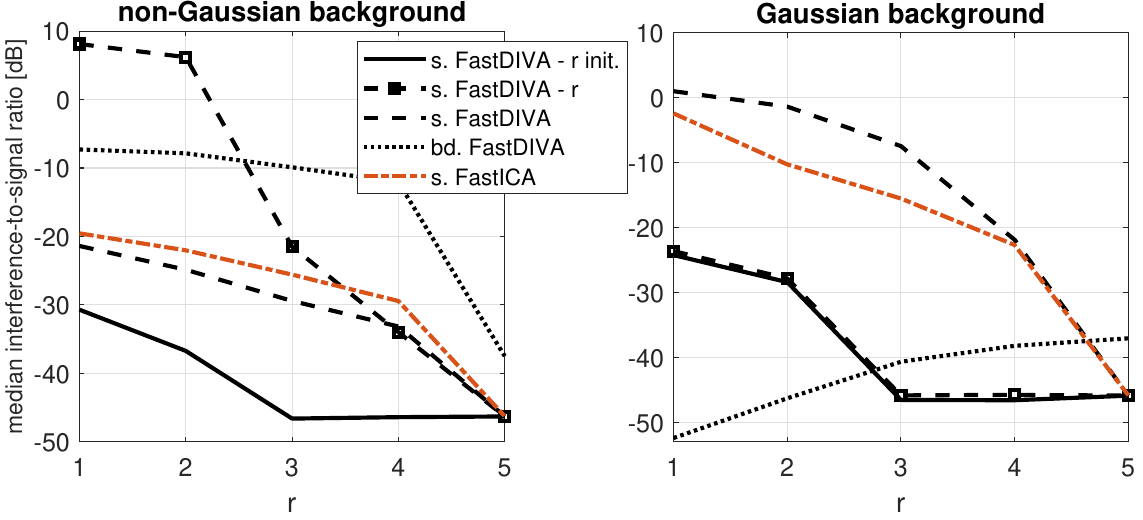}
    \caption{Median ISR of $100$ trials as a function of $r$, $r=1,\dots,d$, $d=5$, achieved by separating dynamic mixtures obeying condition (C1). For $r=d=5$, the mixtures are static; "s." and "bd." stand for symmetric and block-deflation, respectively; "r" means that only $r$ signals are being separated; "init" means a controled initialization.}
    \label{fig:fig3}
\end{figure}

\begin{figure}
    \centering
    \includegraphics[width=0.8\linewidth]{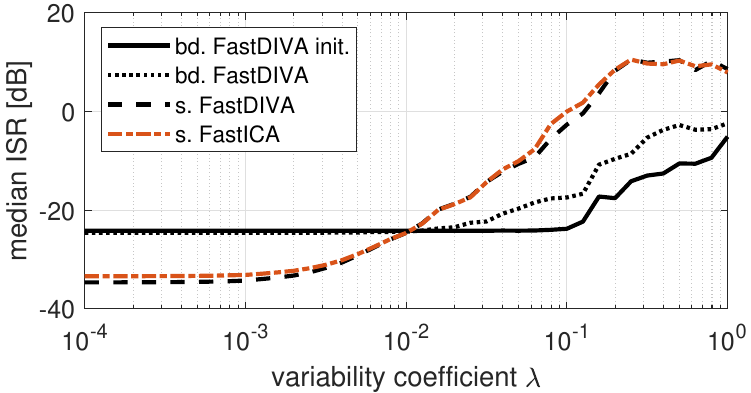}
    \caption{Median ISR as a function of variability coefficient achieved in separation of real-valued mixtures of speech signals from Fig.~\ref{fig:speechsignals} obeying condition (C2).}
    \label{fig:fig4}
\end{figure}

The results of these experiments are shown in Figures~\ref{fig:fig3} and \ref{fig:fig4} in terms of median ISR computed over $100$ trials for each $r$ and $\lambda$, respectively. The median is used instead of the average because, in dynamic settings, the algorithms can fail in many more trials than in the static case; the results indicate that such failures mainly depend on the initializations. In legends, "s." and "bd." are acronyms for the symmetric and block-deflation variants, respectively; "r" means that only $r$ signals are being separated; "init" means that the algorithm is initialized in a vicinity of the correct solution.

Fig.~\ref{fig:fig3} shows that, in both background settings, symmetric FastDIVA yields excellent ISR on the mixtures obeying (C1) provided that it is properly initialized and the true number of signals that obey CSV is known. Without a proper initialization, its performance is close to symmetric FastICA, i.e., when $T=1$ blocks are assumed. In the static case $r=d$, the symmetric algorithms achieve the same superior performance (median ISR about $-45$~dB); it is worth pointing out that, when $r=d$, FastDIVA assumes the overestimated number of blocks ($T=5$), nevertheless, this phenomenon does not deteriorate its performance. Block-deflation FastDIVA performs well in the Gaussian background setting and achieves a lower median ISR when $r=d$ as compared to the symmetric algorithms. The latter observation agrees with the results of previous theoretical analyses of the symmetric and deflation approaches \cite{tichavsky2006} (the symmetric one is usually more accurate; the accuracy of the deflation one depends on the order in which the signals are being separated).

Fig.~\ref{fig:fig4} shows results of the experiment with mixtures (C2), which is, in fact, suitable for block-deflation FastDIVA. This algorithm tends to yield a constant median ISR until $\lambda\approx 10^{-1}$. This is indicative of the fact that the algorithm's performance is equivariant, i.e., independent of the mixing parameters, as are the theoretical bounds \eqref{eq:finalISR} and Cram\'er-Rao bounds in \cite{kautsky2020CRLB}. For higher values of $\lambda$, the probability grows for the algorithm getting stuck in a local extreme; this tendency deteriorates the median ISR. Symmetric FastDIVA and FastICA yield similar median ISR outputs in this scenario. For very small $\lambda$ values, i.e., when the mixture is almost static, they achieve a better ISR than the block-deflation variant, which agrees with the observation shown in Fig.~\ref{fig:fig3} for $r=d$. With growing $\lambda$, the performance of symmetric FastDIVA drops down because the mixture does not meet the condition (C1).

\subsection{Semi-Online Blind Source Extraction}
Here, the application of CSV and one-unit FastDIVA is presented in an online BSE problem where a SOI is being extracted sequentially block-by-block. In such processing, the selection of the length of blocks plays an important role. It affects the key features of the online system: extraction accuracy, adaptability, and susceptibility to the discontinuity problem. 

The benefit of using CSV in online processing is that it allows for dynamics within the block by setting $T>1$. 
With CSV, we can increase the block length without reducing the  time-resolution. We verify this feature in a simulated example where online BSE with $T=1$ and $T>1$ are compared. 

The data are generated as follows. In one trial, a random instantaneous real-valued mixture ($K=1$) of dimension $d=10$ involving one moving laplacean SOI and $8$ static interfering laplacean sources is generated. The sources have zero mean and unit variance. 
The mixing vector related to the SOI is continuously changing in a linear manner so that its value at the $n$th sample, $n=1,\dots,N$, is 
\begin{equation}
    \mathbf{a}_n = \left(1-\frac{n-1}{N-1}\right)\mathbf{a}_1 + \left(\frac{n-1}{N-1}\right)\mathbf{a}_N,
\end{equation}
where $\mathbf{a}_1$ and $\mathbf{a}_N$ are random vectors with unit norm. The movement speed of the SOI is controlled through the angular distance between $\mathbf{a}_1$ and $\mathbf{a}_N$. Gaussian additive noise with zero mean and $0.1$ variance is added to the mixture. The total length of data is $N = 60,000$.

The separating vector is initialized by the LCMP beamformer \cite{vantrees2002} steered in the directions given by $\mathbf{a}_1$ and $\mathbf{a}_N$. The data are then processed block-by-block (with overlap) by performing one one-unit FastDIVA iteration per block, initialized by the separating vector from the previous block. 

The extraction accuracy is evaluated in terms of Signal-to-Interference-plus-Noise Ratio (SINR) and Signal-to-Distortion Ratio (SDR) where the latter is defined as 
\begin{equation}
    {\rm SDR} = \frac{\widehat{\rm E}[\tilde{s}^2]}{\min_\alpha 	\widehat{\rm E}[ ( \tilde{s} - \alpha s  )^2 ]},
\end{equation}
where $\tilde{s}$ is the SOI component within the extracted signal $\widehat{s}$, and $s$ is the true SOI. Table \ref{table:results_online_exp} shows the results averaged over $100$ trials as they depend on the block length, block shift, and the angle between $\mathbf{a}_1$ and $\mathbf{a}_N$, denoted as $\angle({\bf a}_1, {\bf a}_N)$.

\begin{table}
\caption{Results of the simulated online BSE in terms of SINR and SDR 
[dB] averaged over 100 trials.}
\label{table:results_online_exp}
\begin{adjustbox}{width=\columnwidth,center}
\begin{tabular}{|c|c|c||c|c||c|c||c|c|}
\hline
\multicolumn{3}{|c||}{\textbf{\begin{tabular}[c]{@{}c@{}}Angle\\ $\angle({\bf a}_1, {\bf a}_N)$\end{tabular}    }}                                                                                                                     & \multicolumn{2}{c||}{0\degree}   
& \multicolumn{2}{c||}{10\degree}      & \multicolumn{2}{c|}{30\degree}    \\ \hline\hline
\textbf{\begin{tabular}[c]{@{}c@{}}Block\\ length\end{tabular}} & \textbf{\begin{tabular}[c]{@{}c@{}}Block\\ shift\end{tabular}} & \multicolumn{1}{c||}{\textbf{\begin{tabular}[c]{@{}c@{}}\#CSV \\ blocks \\ ($T$)\end{tabular}}} & \textbf{SINR} & \textbf{SDR} & 
\textbf{SINR} & \textbf{SDR} & \textbf{SINR} & \textbf{SDR}
\\ 
\hline
500 & 100 & 1 & 13.8 & 25.2 & 
11.3 & 21.3 & 6.3 & 9.0  \\
\hline
2500 & 500 & 1 & 21.2 & 32.8 &
8.0 & 11.4 & 7.5 & 3.6 \\ 
\hline
5000 & 1000 & 1 & 24.5 & 36.4 &
8.2 & 6.5 & 13.0 & 2.4  \\
\hline \hline
500 & 100 & 5 & 13.9 & 25.9 &
12.8 & 23.5 & 12.9 & 20.0 \\
\hline
2500 & 500 & 5 & 21.2 & 32.8 & 
18.4 & 24.8 & 19.7 & 13.9 \\
\hline
5000 & 1000 & 5 & 24.3 & 36.4 & 
20.4 & 25.0 & 23.0 & 17.6  \\ \hline
\end{tabular}
\end{adjustbox}
\end{table}

When $\angle({\bf a}_1, {\bf a}_N)=0$, the mixture is static. Here, the SINR and SDR are obviously increasing with the growing block length. As expected, $T=5$ brings no advantage compared to $T=1$, in this case. By contrast, when the SOI is moving and $\angle({\bf a}_1, {\bf a}_N)>0$, the processing with $T=5$ brings significantly better SINR as well as SDR compared to $T=1$. 

Also, by detailed inspection of the values of SDR when $\angle({\bf a}_1, {\bf a}_N)=30 \degree$, we can see that the optimum block length is different for $T=5$ than for $T=1$.

\subsection{Blind speech extraction of a moving speaker: a case study}
\begin{figure}
 \centering
 \includegraphics[width=0.99\linewidth]{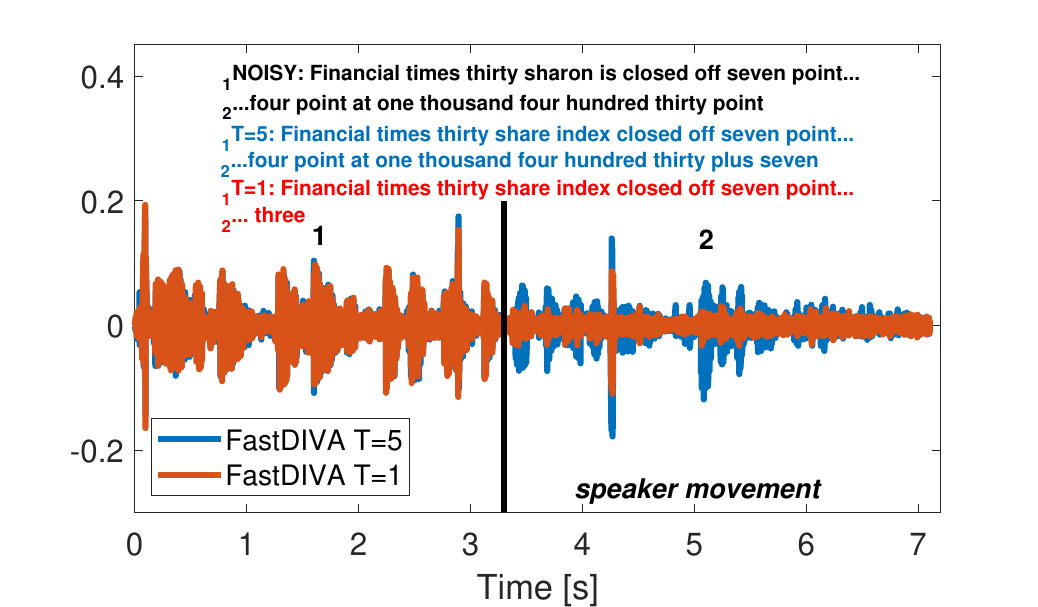}
 \caption{\label{fig:casestudy} Extracted utterances from a noisy recording of a moving speaker. The ground truth transcription is ``Financial times thirty share index closed off seven point four points at one thousand four hundred thirty point seven''; NOISY stands for the automatic transcription from the original noisy recording by the first microphone; $T=1$ and $T=5$ correspond to the One-unit FastDIVA outputs considering, respectively, the static and the CSV mixing model.}
\end{figure}

To demonstrate the applicability of the proposed method, we consider a  speech enhancement task where the speaker is moving. A six-channel recording of a speaker uttering in a multi-source noisy environment is taken from the CHiME-4 challenge database\footnote{The presented utterance is {\tt F04\_053C010W\_BUS.WAV}.} \cite{chime_data_soft}. One-unit FastDIVA is applied in the short-term Fourier transform domain (the window length is 512 samples and the hop size is 128) in order to extract the speech. Fig.~\ref{fig:casestudy} shows the resulting signals and compares the ground truth transcription with automatic transcriptions by the Google Speech-to-Text system\footnote{The transcriptions were performed by the system on December 8, 2020 at {\tt https://cloud.google.com/speech-to-text}.}.

Within this particular recording, the speaker moves out of its initial position in the interval $3.3-7.1$~s. One-unit FastDIVA with $T=1$ (static mixing model) focuses only on the initial speaker position. This causes that the extracted voice is vanishing during the interval of the movement; the corresponding part of the automatic transcription is therefore erroneous. When the algorithm is used with $T=5$, the whole utterance is successfully extracted, which results in a significantly more accurate transcription.

\section{Conclusions}
In this paper, we propose powerful BSS algorithms suitable for separating dynamic CSV-separable mixtures where the avoidance of the discontinuity problem is guaranteed. Joint separation similar to IVA, which helps us solve the permutation ambiguity, is considered as well. One-unit FastDIVA has been shown as effective for the BSE when the SOI obeys the CSV mixing model. The performance analysis has been derived for a general model pdf of the SOI, and it has been proven that One-unit FastDIVA attains the Cram\'er-Rao lower bound asymptotically when the background is circular Gaussian and the model pdf corresponds to the true one. Symmetric and block-deflation FastDIVA have been validated in the problem of separating several signals from CSV-separable mixtures obeying condition (C1) and (C2), respectively. The results of experiments indicate that the algorithms achieve superior interference-to-signal ratio compared to methods assuming the conventional static mixing model, especially, when the time-variability of the mixture is mild. The reliability of the separation can be supported by a proper initialization; other forms of partial knowledge about the mixing parameters might be considered in future works, as in \cite{brendel2020}. 

By generalizing the mixing model, we have touched on the very basis of the problem that ICA and IVA solve. For this reason, many theoretical and practical questions arise. In particular, the question is what mixtures can be approximated well enough by a CSV-separable model with $T\ll N$. Equivalently: Which of the signals in \eqref{eq:mixingmodel} can be extracted based on the CSV model and what does this mean in practice? The experiments here and elsewhere \cite{koldovsky2019icassp,jansky2020} show that the mixture can be dynamic only to a limited extent. For example, it is better when only some sources are moving. In addition, their movements (within the processed batch of data) should be spatially limited so that a separating vector that covers the entire motion space exists. That space should not be intersected by the motion trajectories of the other sources. A more specific analysis is the subject of further research.

Finally, it worth pointing to the fact that the static ICA/IVA problem has the property that if at most one of the sources is Gaussian and the others are non-Gaussian, then there are no false solutions. That is, there are no independent components that do not correspond to the original signals except for the scale and order \cite{theis2004}. For $T>1$, similar analysis does not exist yet.

\section*{Appendix A: Proof of Proposition~1}
By considering $N=+\infty$, all estimated values and averages are replaced by the true expectation values. Using the complex derivative identities \cite{kreutzdelgado2009}, it holds that $\mathcal{C}_{\rm OG}$ is a real function, so $\frac{\partial}{\partial{\bf w}^H}\mathcal{C}_{\rm OG}=(\frac{\partial}{\partial{\bf w}^T}\mathcal{C}_{\rm OG})^*$, and by definition \eqref{eq:normalizedgrad}
\begin{align}
{\bf H}_1&=\frac{\partial^2\mathcal{C}_{\rm OG}}{\partial{\bf w}^T\partial{\bf w}}=
\frac{\partial \nabla^H}{\partial{\bf w}}=\left[\frac{\partial}{\partial{\bf w}^H}\left({\bf a}^T-\nu^{-1}{\rm E}\left[\phi\frac{{\bf x}^T}{\sigma}\right]\right)\right]^*\label{A:H1}\\
{\bf H}_2&=\frac{\partial^2\mathcal{C}_{\rm OG}}{\partial{\bf w}^H\partial{\bf w}}=
\frac{\partial \nabla^T}{\partial{\bf w}}=\frac{\partial}{\partial{\bf w}}\left({\bf a}^T-
\nu^{-1}{\rm E}\left[\phi\frac{{\bf x}^T}{\sigma}\right]\right)\label{A:H2},
\end{align}
where $\phi(\frac{s}{\sigma})$ is, for brevity, written without the argument (which is always $\frac{s}{\sigma}$). Note that the dependent variables on ${\bf w}$ are $s={\bf w}^H{\bf x}$, ${\bf a}$ through the OGC, and $\sigma$ through \eqref{eq:sigma}; $\nu$ is treated as a constant in \eqref{A:H1} and \eqref{A:H2}.  Using the following auxiliary expressions,
\begin{align}
 \frac{\partial {\bf a}^T}{\partial{\bf w}}&=
 \frac{{\bf C}^*}{\sigma^2}-{\bf a}^*{\bf a}^T &
 \frac{\partial {\bf a}^T}{\partial{\bf w}^H}&=
 -{\bf a}{\bf a}^T,\label{A:exp2}\\
 \frac{\partial}{\partial{\bf w}}\frac{1}{\sigma}&=
 -\frac{{\bf a}^*}{2\sigma} &
 \frac{\partial}{\partial{\bf w}^H}\frac{1}{\sigma}&=
 -\frac{{\bf a}}{2\sigma},\\
 \frac{\partial}{\partial{\bf w}}\frac{s^*}{\sigma}&=
 \frac{{\bf x}^*}{\sigma}-\frac{s^*{\bf a}^*}{2\sigma} &
 \frac{\partial}{\partial{\bf w}^H}\frac{s}{\sigma}&=
 \frac{{\bf x}}{\sigma}-\frac{s{\bf a}}{2\sigma},
\end{align}
straightforward computations give
\begin{align}
&\frac{\partial}{\partial{\bf w}^H}\phi\frac{{\bf x}^T}{\sigma}=
   \frac{\partial\phi}{\partial s}\left(\frac{{\bf x}}{\sigma}-\frac{s{\bf a}}{2\sigma}\right)\frac{{\bf x}^T}{\sigma}-
   \frac{\partial\phi}{\partial s^*}
   \frac{s^*{\bf a}}{2\sigma}
   \frac{{\bf x}^T}{\sigma}
   -\phi\frac{{\bf a}{\bf x}^T}{2\sigma},\nonumber\\
   &\frac{\partial}{\partial{\bf w}}\phi\frac{{\bf x}^T}{\sigma}=
   \frac{\partial\phi}{\partial s^*}\left(\frac{{\bf x}^*}{\sigma}-\frac{s^*{\bf a}^*}{2\sigma}\right)\frac{{\bf x}^T}{\sigma}-\frac{\partial\phi}{\partial s}\frac{s{\bf a}^*}{2\sigma}\frac{{\bf x}^T}{\sigma}
      -\phi\frac{{\bf a}^*{\bf x}^T}{2\sigma}.\nonumber
\end{align}
By taking the expectation values of the latter expressions and using the fact that ${\bf x}={\bf a}s+{\bf y}$ where ${\bf s}$ and ${\bf y}$ have zero mean values and are independent, we obtain
\begin{align}
\frac{\partial}{\partial{\bf w}^H}{\rm E}\left[\phi\frac{{\bf x}^T}{\sigma}\right]&=
    \frac{1}{2}\left(\eta-\xi-\nu\right){\bf a}{\bf a}^T+\frac{\rho}{\sigma^2}{\bf P}_{\bf y},\label{A:exp4}\\
    \frac{\partial}{\partial{\bf w}}{\rm E}\left[\phi\frac{{\bf x}^T}{\sigma}\right]&=
    \frac{1}{2}\left(\xi-\eta-\nu\right){\bf a}^*{\bf a}^T+\frac{\rho}{\sigma^2}{\bf C}_{\bf y}^*\label{A:exp5},
\end{align}
where ${\bf P}_{\bf y}={\rm E}[{\bf y}{\bf y}^T]$ is the pseudo-covariance of ${\bf y}$, which is zero due to the assumption of circularity of the background signals. Putting \eqref{A:exp4} with ${\bf P}_{\bf y}={\bf 0}$ and \eqref{A:exp2} into \eqref{A:H1}, we get \eqref{eq:H1}.

Finally, note that
${\bf C}={\bf a}{\bf a}^H\sigma^2+{\bf C}_{\bf y}$, so \eqref{A:exp5} is equal to $\frac{\partial}{\partial{\bf w}}{\rm E}\left[\phi\frac{{\bf x}^T}{\sigma}\right]=(\nu c_3-\rho){\bf a}^*{\bf a}^T+\frac{\rho}{\sigma^2}{\bf C}^*$, where $c_3$ is defined by \eqref{eq:c3}. Putting this and \eqref{A:exp2} into \eqref{A:H2}, and using definitions \eqref{eq:c1} and \eqref{eq:c2}, we get \eqref{eq:H2}.\hfill\rule{1.2ex}{1.2ex}

\section*{Appendix B: Proof of Proposition~2}
Applying the Woodbury identity to (\ref{eq:H2}) gives 
\begin{equation}
    {\bf H}_2^{-1}=
    \frac{1}{c_1}\left({\bf C}^{-1}-{\bf C}^{-1}{\bf a}\Bigl(\frac{c_1}{c_2}+{\bf a}^H{\bf C}{\bf a}\Bigr)^{-1}{\bf a}^H{\bf C}^{-1}\right)^T.\label{B:exp1}
\end{equation}
Using the following equalities due to the OGC imposed between ${\bf a}$ and ${\bf w}$,
\begin{align}
    {\bf C}{\bf w}({\bf a}^H{\bf C}^{-1}{\bf a})&={\bf a}, &
    {\bf w}^H{\bf a}&=1,\label{B:exp3}\\
    {\bf a}^H{\bf C}^{-1}{\bf a}&=\frac{1}{\sigma^2}, &
    {\bf C}^{-1}{\bf a} &= \frac{\bf w}{\sigma^2},\label{B:exp4}
\end{align}
\eqref{B:exp1} can be written as
\begin{equation}
  {\bf H}_2^{-1}=\frac{1}{c_1}\left({\bf C}^{-1}-\frac{c_2}{\sigma^2(\sigma^2c_1+c_2)}{\bf w}{\bf w}^H
  \right)^T,
\end{equation}
and since ${\bf H}_1^{*}=c_3{\bf a}{\bf a}^T$, after simplifications, 
\begin{align}
  {\bf H}_1^{*}{\bf H}_2^{-1} &=  -{\bf a}{\bf w}^T\label{B:exp2}\\
  {\bf H}_1^{*}{\bf H}_2^{-1}{\bf H}_1 &=  -c_3^*{\bf a}{\bf a}^H\\
  {\bf H}_2^*-{\bf H}_1^*{\bf H}_2^{-1}{\bf H}_1 &= 
  c_1^*({\bf C}-\sigma^2{\bf a}{\bf a}^H),
\end{align}
from which, by the latter equation, \eqref{eq:H} follows.

Next, we show that $\hat{\bf H}_1^*\hat{\bf H}_2^{-1}{\nabla}^*={\bf 0}$. Let us denote 
\begin{equation}
    {\bf f}=\hat{\rm E}\left[\phi\left(\frac{{\bf w}^H{\bf x}}{\sigma}\right)\frac{\bf x}{\sigma}\right]\label{B:f},
\end{equation}
which is the expression that appears in \eqref{eq:normalizedgrad}; so we can write that $\nabla={\bf a}-\hat\nu^{-1}{\bf f}$. From the definition of $\hat\nu$, it follows that ${\bf w}^H{\bf f}=\hat\nu$, and by using \eqref{B:exp2}, $\hat{\bf H}_1^*\hat{\bf H}_2^{-1}{\nabla}^*={\bf 0}$.
Thus, we receive the update (40).

Let 
\begin{equation}\label{eq:He}
    \hat{\bf H}_\epsilon=\left(\frac{\hat\nu-\hat\rho}{\hat\nu}\right)^*\left(\frac{\widehat{\bf C}}{\hat\sigma^2}-\epsilon{\bf a}{\bf a}^H\right),
\end{equation}
so that $\lim_{\epsilon\rightarrow 1}\hat{\bf H}_\epsilon=\hat{\bf H}$, cf. (\ref{eq:H}).
Using the Woodbury identity, \eqref{B:exp3}, and \eqref{B:exp4}, we get
\begin{equation}
\hat{\bf H}_\epsilon^{-1}=\left(\frac{\hat\nu}{\hat\nu-\hat\rho}\right)^*   
\left(\hat\sigma^2\widehat{\bf C}^{-1}+\frac{\epsilon}{\hat\sigma^2(1-\epsilon)}{\bf w}{\bf w}^H\right).
\end{equation}
Using \eqref{B:f}, $\nabla={\bf a}-\hat\nu^{-1}{\bf f}$, ${\bf w}^H{\bf f}=\hat\nu$, and ${\bf w}^H\nabla=0$. Thus
\begin{equation}
 \hat{\bf H}_\epsilon^{-1}\nabla= \left(\frac{\hat\nu}{\hat\nu-\hat\rho}\right)^*   
\hat\sigma^2\widehat{\bf C}^{-1}\nabla~.
\end{equation}
The update (\ref{eq:newtonupdatefinal}) readily follows.
\hfill\rule{1.2ex}{1.2ex}



\begin{thebibliography}{10}

\bibitem{herault1987}
J.~Herault and C.~Jutten, ``Space or time adaptive signal processing by neural
  network models,'' in {\em AIP Conference Proceedings 151 on Neural Networks
  for Computing}, (Woodbury, NY, USA), pp.~206--211, American Institute of
  Physics Inc., 1987.

\bibitem{comon1994}
P.~Comon, ``Independent component analysis, a new concept?,'' {\em Signal
  Processing}, vol.~36, pp.~287--314, 1994.

\bibitem{cardoso1998}
J.~F. Cardoso, ``Blind signal separation: statistical principles,'' {\em
  Proceedings of the {IEEE}}, vol.~86, pp.~2009--2025, Oct 1998.

\bibitem{lee1998}
T.-W. Lee, {\em Independent Component Analysis - Theory and Applications}.
\newblock Kluwer Academic Publishers, 1998.

\bibitem{hyvarinen2001}
A.~Hyv\"{a}rinen, J.~Karhunen, and E.~Oja, {\em Independent Component
  Analysis}.
\newblock John Wiley \& Sons, 2001.

\bibitem{cichocki2002}
A.~Cichocki and S.~Amari, {\em Adaptive Blind Signal and Image Processing}.
\newblock John Wiley \& Sons, 2002.

\bibitem{comon2010handbook}
P.~Comon and C.~Jutten, {\em Handbook of Blind Source Separation: Independent
  Component Analysis and Applications}.
\newblock Independent Component Analysis and Applications Series, Elsevier
  Science, 2010.

\bibitem{adali2014b}
T.~{Adal\i}, C.~{Jutten}, A.~{Yeredor}, A.~{Cichocki}, and E.~{Moreau},
  ``Source separation and applications [from the guest editors],'' {\em IEEE
  Signal Processing Magazine}, vol.~31, no.~3, pp.~16--17, 2014.

\bibitem{kitamura2018}
D.~Kitamura, S.~Mogami, Y.~Mitsui, N.~Takamune, H.~Saruwatari, N.~Ono,
  Y.~Takahashi, and K.~Kondo, ``Generalized independent low-rank matrix
  analysis using heavy-tailed distributions for blind source separation,'' {\em
  EURASIP Journal on Advances in Signal Processing}, vol.~2018, p.~28, May
  2018.

\bibitem{brendel2020}
A.~{Brendel}, T.~{Haubner}, and W.~{Kellermann}, ``A unified probabilistic view
  on spatially informed source separation and extraction based on independent
  vector analysis,'' {\em IEEE Transactions on Signal Processing}, vol.~68,
  pp.~3545--3558, 2020.

\bibitem{haykin2010}
W.~Liu, J.~C. Principe, and S.~Haykin, {\em Kernel Adaptive Filtering: A
  Comprehensive Introduction}.
\newblock Wiley Publishing, 1st~ed., 2010.

\bibitem{welling2004}
M.~{Welling}, R.~S. {Zemel}, and G.~E. {Hinton}, ``Probabilistic sequential
  independent components analysis,'' {\em IEEE Transactions on Neural
  Networks}, vol.~15, no.~4, pp.~838--849, 2004.

\bibitem{taniguchi2014}
T.~{Taniguchi}, N.~{Ono}, A.~{Kawamura}, and S.~{Sagayama}, ``An
  auxiliary-function approach to online independent vector analysis for
  real-time blind source separation,'' in {\em HSCMA 2014}, pp.~107--111, May
  2014.

\bibitem{hsu2016}
S.~H. {Hsu}, T.~R. {Mullen}, T.~P. {Jung}, and G.~{Cauwenberghs}, ``Real-time
  adaptive eeg source separation using online recursive independent component
  analysis,'' {\em IEEE Transactions on Neural Systems and Rehabilitation
  Engineering}, vol.~24, no.~3, pp.~309--319, 2016.

\bibitem{chien2013}
J.~{Chien} and H.~{Hsieh}, ``Nonstationary source separation using sequential
  and variational bayesian learning,'' {\em IEEE Transactions on Neural
  Networks and Learning Systems}, vol.~24, no.~5, pp.~681--694, 2013.

\bibitem{amari1996}
S.~Amari, A.~Cichocki, and H.~H. Yang, ``A new learning algorithm for blind
  signal separation,'' in {\em Proceedings of Neural Information Processing
  Systems}, pp.~757--763, 1996.

\bibitem{laheld1996}
J.~F. Cardoso and B.~H. Laheld, ``Equivariant adaptive source separation,''
  {\em {IEEE} Transactions on Signal Processing}, vol.~44, pp.~3017--3030, Dec
  1996.

\bibitem{mukai2005}
R.~Mukai, H.~Sawada, S.~Araki, and S.~Makino, {\em Real-Time Blind Source
  Separation for Moving Speech Signals}, pp.~353--369.
\newblock Berlin, Heidelberg: Springer Berlin Heidelberg, 2005.

\bibitem{nesta2011taslp}
F.~{Nesta}, T.~S. {Wada}, and B.~{Juang}, ``Batch-online semi-blind source
  separation applied to multi-channel acoustic echo cancellation,'' {\em IEEE
  Transactions on Audio, Speech, and Language Processing}, vol.~19, no.~3,
  pp.~583--599, 2011.

\bibitem{akhtar2012}
M.~T. {Akhtar}, T.~{Jung}, S.~{Makeig}, and G.~{Cauwenberghs}, ``Recursive
  independent component analysis for online blind source separation,'' in {\em
  2012 IEEE International Symposium on Circuits and Systems (ISCAS)},
  pp.~2813--2816, 2012.

\bibitem{khan2015}
A.~H. Khan, M.~Taseska, and E.~A.~P. Habets, {\em A Geometrically Constrained
  Independent Vector Analysis Algorithm for Online Source Extraction},
  pp.~396--403.
\newblock Cham: Springer International Publishing, 2015.

\bibitem{hsu2014}
S.~{Hsu}, T.~{Mullen}, T.~{Jung}, and G.~{Cauwenberghs}, ``Online recursive
  independent component analysis for real-time source separation of
  high-density eeg,'' in {\em 2014 36th Annual International Conference of the
  IEEE Engineering in Medicine and Biology Society}, pp.~3845--3848, 2014.

\bibitem{koldovsky2019icassp}
Z.~Koldovsk\'y, J.~M\'alek, and J.~Jansk\'y, ``Extraction of independent vector
  component from underdetermined mixtures through block-wise determined
  modeling,'' in {\em Proceedings of {IEEE} International Conference on Audio,
  Speech and Signal Processing}, vol.~7903--7907, May 2019.

\bibitem{jansky2020}
J.~Jansk\'y, Z.~Koldovsk\'y, J.~M\'alek, T.~Kounovsk\'y, and J.~\v{C}mejla,
  ``Fast algorithm for blind independence-based extraction of a moving
  speaker,'' {\em arXiv}, 2020, 2002.12619.

\bibitem{kautsky2020CRLB}
V.~{Kautsk\'y}, Z.~{Koldovsk\'y}, P.~{Tichavsk\'y}, and V.~{Zarzoso},
  ``{Cram\'er-Rao} bounds for complex-valued independent component extraction:
  Determined and piecewise determined mixing models,'' {\em IEEE Transactions
  on Signal Processing}, vol.~68, pp.~5230--5243, 2020.

\bibitem{hyvarinen1999}
A.~Hyv\"{a}rinen, ``Fast and robust fixed-point algorithm for independent
  component analysis,'' {\em {IEEE} Transactions on Neural Networks}, vol.~10,
  no.~3, pp.~626--634, 1999.

\bibitem{lee2007fast}
I.~Lee, T.~Kim, and T.-W. Lee, ``Fast fixed-point independent vector analysis
  algorithms for convolutive blind source separation,'' {\em Signal
  Processing}, vol.~87, no.~8, pp.~1859--1871, 2007.

\bibitem{lee1997}
T.-W. Lee, A.~Bell, and R.~Orglmeister, ``Blind source separation of real world
  signals,'' in {\em Proc. ICNN}, pp.~2129--2135, June 1997.

\bibitem{adali2015}
T.~Adal\i, Y.~Levin-Schwartz, and V.~D. Calhoun, ``Multimodal data fusion using
  source separation: Two effective models based on ica and iva and their
  properties,'' {\em Proceedings of the IEEE}, vol.~103, pp.~1478--1493, Sep.
  2015.

\bibitem{lahat2016}
D.~Lahat and C.~Jutten, ``Joint independent subspace analysis using
  second-order statistics,'' {\em IEEE Transactions on Signal Processing},
  vol.~64, pp.~4891--4904, Sept 2016.

\bibitem{chen2016}
X.~{Chen}, Z.~J. {Wang}, and M.~{McKeown}, ``Joint blind source separation for
  neurophysiological data analysis: Multiset and multimodal methods,'' {\em
  IEEE Signal Processing Magazine}, vol.~33, pp.~86--107, May 2016.

\bibitem{weiss2018}
A.~Weiss, S.~A. Cheema, M.~Haardt, and A.~Yeredor, ``Performance analysis of
  the gaussian quasi-maximum likelihood approach for independent vector
  analysis,'' {\em IEEE Transactions on Signal Processing}, vol.~66,
  pp.~5000--5013, Oct 2018.

\bibitem{sawada2004sap}
H.~Sawada, R.~Mukai, S.~Araki, and S.~Makino, ``A robust and precise method for
  solving the permutation problem of frequency-domain blind source
  separation,'' {\em {IEEE} Transactions on Speech and Audio Processing},
  vol.~12, pp.~530--538, Sept. 2004.

\bibitem{kim2006}
T.~Kim, I.~Lee, and T.~Lee, ``Independent vector analysis: Definition and
  algorithms,'' in {\em 2006 Fortieth Asilomar Conference on Signals, Systems
  and Computers}, pp.~1393--1396, Oct 2006.

\bibitem{koldovsky2019TSP}
Z.~Koldovsk\'y and P.~Tichavsk\'y, ``Gradient algorithms for complex
  non-gaussian independent component/vector extraction, question of
  convergence,'' {\em IEEE Transactions on Signal Processing}, vol.~67,
  pp.~1050--1064, Feb 2019.

\bibitem{gannot2001}
S.~Gannot, D.~Burshtein, and E.~Weinstein, ``Signal enhancement using
  beamforming and nonstationarity with applications to speech,'' {\em IEEE
  Transactions on Signal Processing}, vol.~49, pp.~1614--1626, Aug 2001.

\bibitem{kautsky2017}
V.~Kautsk\'y, Z.~Koldovsk\'y, and P.~Tichavsk\'y, ``{Cram\'er-Rao}-induced
  bound for interference-to-signal ratio achievable through non-gaussian
  independent component extraction,'' in {\em 2017 IEEE International Workshop
  on Computational Advances in Multi-Sensor Adaptive Processing (CAMSAP)},
  pp.~94--97, Dec 2017.

\bibitem{pham2001b}
D.-T.~A. Pham, ``Contrast functions for blind separation and deconvolution of
  sources,'' in {\em Proceedings of International Conference on Independent
  Component Analysis and Signal Separation}, Dec 2001.

\bibitem{pham2006}
D.-T.~A. Pham, ``Blind partial separation of instantaneous mixtures of
  sources,'' in {\em Proceedings of International Conference on Independent
  Component Analysis and Signal Separation}, pp.~868--875, Springer Berlin
  Heidelberg, 2006.

\bibitem{vantrees2002}
H.~L. Van~Trees, {\em Optimum Array Processing: Part IV of Detection,
  Estimation, and Modulation Theory}.
\newblock John Wiley \& Sons, Inc., 2002.

\bibitem{koldovsky2017eusipco}
Z.~Koldovsk\'{y}, P.~Tichavsk\'y, and V.~Kautsk\'y, ``Orthogonally constrained
  independent component extraction: Blind {MPDR} beamforming,'' in {\em
  Proceedings of European Signal Processing Conference}, pp.~1195--1199, Sept.
  2017.

\bibitem{li2008}
H.~Li and T.~Adal{\i}, ``Complex-valued adaptive signal processing using
  nonlinear functions,'' {\em EURASIP Journal on Advances in Signal
  Processing}, vol.~2008, p.~765615, Feb 2008.

\bibitem{huber1985}
P.~J. Huber, ``Projection pursuit,'' {\em Ann. Statist.}, vol.~13,
  pp.~435--475, June 1985.

\bibitem{bingham2000}
E.~Bingham and A.~Hyv\"{a}rinen, ``A fast fixed-point algorithm for independent
  component analysis of complex valued signals,'' {\em International Journal of
  Neural Systems}, vol.~10, pp.~1--8, Feb. 2000.

\bibitem{kreutzdelgado2009}
K.~Kreutz-Delgado, ``The complex gradient operator and the cr-calculus,'' {\em
  arXiv}, 2009, 0906.4835.

\bibitem{delfosse1995}
N.~Delfosse and P.~Loubaton, ``Adaptive blind separation of independent
  sources: A deflation approach,'' {\em Signal Processing}, vol.~45, no.~1,
  pp.~59 -- 83, 1995.

\bibitem{cardoso1994}
J.-F. Cardoso, ``On the performance of orthogonal source separation
  algorithms,'' in {\em Proceedings of European Signal Processing Conference},
  pp.~776--779, Sept. 1994.

\bibitem{Porat2008}
B.~Porat, {\em Digital Processing of Random Signals: Theory and Methods}.
\newblock Dover Publications, 2008.

\bibitem{hyvarinen1997b}
A.~Hyv\"{a}rinen, ``One-unit contrast functions for independent component
  analysis: a statistical analysis,'' in {\em Neural Networks for Signal
  Processing VII. Proceedings of the 1997 IEEE Signal Processing Society
  Workshop}, pp.~388--397, Sep 1997.

\bibitem{tichavsky2006}
P.~Tichavsk\'y, Z.~Koldovsk\'y, and E.~Oja, ``Performance analysis of the
  fastica algorithm and {Cram\'er-Rao} bounds for linear independent component
  analysis,'' {\em {IEEE} Transactions on Signal Processing}, vol.~54,
  pp.~1189--1203, April 2006.

\bibitem{koldovsky2019quick}
Z.~Koldovsk\'y and V.~Kautsk\'y, ``Quick algorithms for independent vector
  extraction and analysis based on exact newton-raphson optimization,'' {\em
  arXiv}, 2019, 1910.10242.

\bibitem{rati2007}
P.~Tichavsk{\'y}, Z.~Koldovsk{\'y}, and E.~Oja, {\em Speed and Accuracy
  Enhancement of Linear ICA Techniques Using Rational Nonlinear Functions},
  pp.~285--292.
\newblock Berlin, Heidelberg: Springer Berlin Heidelberg, 2007.

\bibitem{loeschCRB}
B.~Loesch and B.~Yang, ``{Cram\' er--Rao} bound for circular and noncircular
  complex independent component analysis,'' in {\em IEEE Trans. Signal
  Processing}, vol.~61, pp.~365--379, Jan 2013.

\bibitem{kautsky2019icassp}
V.~Kautsk\'y, Z.~Koldovsk\'y, and P.~Tichavsk\'y, ``Performance bound for blind
  extraction of non-{Gaussian} complex-valued vector component from {Gaussian}
  background,'' in {\em Proceedings of {IEEE} International Conference on
  Audio, Speech and Signal Processing}, vol.~5287--5291, May 2019.

\bibitem{chime_data_soft}
E.~Vincent, S.~Watanabe, A.~A. Nugraha, J.~Barker, and R.~Marxer, ``An analysis
  of environment, microphone and data simulation mismatches in robust speech
  recognition,'' {\em Computer Speech \& Language}, 2016.

\bibitem{theis2004}
F.~J. {Theis}, ``Uniqueness of real and complex linear independent component
  analysis revisited,'' in {\em 2004 12th European Signal Processing
  Conference}, pp.~1705--1708, 2004.

\end{thebibliography}

\end{document}